\documentclass[11pt]{article}
\usepackage{amsmath,amssymb,color,graphics,epsfig,cite}

\usepackage{amsfonts}

\newcommand{\hoch}[1]{$\, ^{#1}$}

\newcommand{\be}{\begin{equation}}
\newcommand{\ee}{\end{equation}}
\newcommand{\bea}{\setlength\arraycolsep{2pt} \begin{eqnarray}}
\newcommand{\eea}{\end{eqnarray}}
\newcommand{\nn}{\nonumber}

\def\ft#1#2{{\textstyle{\frac{\scriptstyle #1}{\scriptstyle #2} } }}
\def\fft#1#2{{\frac{#1}{#2}}}

\def\0{{\sst{(0)}}}
\def\1{{\sst{(1)}}}
\def\2{{\sst{(2)}}}
\def\3{{\sst{(3)}}}
\def\4{{\sst{(4)}}}
\def\5{{\sst{(5)}}}
\def\6{{\sst{(6)}}}
\def\7{{\sst{(7)}}}
\def\8{{\sst{(8)}}}
\def\sst#1{{\scriptscriptstyle #1}}

\begin{document}

\begin{center}
{\Large {\bf Desclarizing the Wormhole to Black Hole with Negative Mass  }}

\vspace{20pt}

Ze Li\hoch{1}, Hai-Shan Liu\hoch{1} and H. L\"{u}\hoch{1,2}

\vspace{10pt}

{\it \hoch{1}Center for Joint Quantum Studies, Department of Physics,\\
School of Science, Tianjin University, Tianjin 300350, China }

\medskip
		
{\it \hoch{2}The International Joint Institute of Tianjin University, Fuzhou,\\ Tianjin University, Tianjin 300350, China}

\vspace{40pt}

\underline{ABSTRACT}
\end{center}

We construct new black holes using the curvature-induced scalarization mechanism in Einstein gravity coupled to a massless phantom scalar. We find that the general wormhole solutions with independent scalar charge become descalarized so that the resulting black hole scalar hair becomes secondary and a function of the mass only. The long-range force between the two identical black holes can be attractive, zero or repulsive, depending on the mass/scalar charge relation. Furthermore, the black hole mass can be negative. Our finding suggests that exotic matter responsible for wormholes can also lead to exotic black holes.



\thispagestyle{empty}
\pagebreak



\section{Introduction}

In the framework of Einstein's theory of General Relativity, the concept of wormhole is as old as black hole. The Schwarzschild metric was constructed in 1916, but its interpretation as a black hole was acquired much later. In the same year, the wormhole concept was proposed \cite{flamm}. However, it turns out that any traversable wormhole requires exotic matter whose energy-momentum tensor violates the null energy condition, see, e.g.~\cite{Morris:1988cz}. This no-go theorem put a dent in the research enthusiasm on wormholes compared to black holes, even if the research on the topic has never waned.

The recent development in high precision cosmology and the difficulty in its explanation within Einstein's theory propelled the discussion of phantom fields, e.g.~\cite{Caldwell:1999ew}.  Furthermore, the instability associated with a phantom field is not incurable \cite{Piazza:2004df}. The progress of the AdS/CFT correspondence also has made the wormhole more legit \cite{Gao:2016bin}. (See also, e.g., \cite{Liu:2025foo,Kawamoto:2025oko}.) In this paper, we study the black hole implications of the phantom matter that is responsible for the wormhole construction. There exist exact solutions describing the most general spherically-symmetric and static spacetime in Einstein gravity minimally coupled to a free massless scalar. The solutions contain two independent nontrivial parameters, the mass $M$ and the scalar hair or charge $\Sigma$, associated with the asymptotic large-$r$ expansions
\be
g_{tt} \sim -1 + \fft{2M}{r} + \cdots\,,\qquad \phi = \fft{2\Sigma}{r} + \cdots\,.
\ee
However, the general solution cannot be a black hole, since the no-hair theorem excludes the possibility for a black hole to carry such a scalar charge $\Sigma$. For the standard free massless scalar, the solution has a naked singularity \cite{Janis:1968zz}. The spacetime becomes regular and geodesic complete for a phantom massless scalar, where the kinetic term of the scalar has the opposite sign to the usual scalar. The Ellis wormhole is perhaps the simplest traversable wormhole \cite{Ellis:1973yv}, constructed from the free massless phantom scalar, with $M=0$ and nonzero phantom hair $\Sigma$. It is a symmetric wormhole connecting two identical Minkowski spacetimes that can be identified as one. The Ellis wormhole is generalized to contain two independent parameters $(M,\Sigma)$ with $\Sigma>M$ \cite{Bronnikov:1973fh}. The global structure of the general Ellis-Bronnikov wormhole was analysed  and it was shown that it connects two asymmetric Minkowski spacetimes that cannot be transformed from one to the other by a transitive action of the Lorentz group \cite{Huang:2020qmn}. Exact solutions of wormholes can be found in a variety of constructions, e.g., \cite{Goulart:2017iko,Sahoo:2018kct,Maldacena:2018gjk,Carvente:2019gkd,
Chew:2019lsa,Garattini:2019ivd,Huang:2019arj}.

One of the important distinctions between a regular massless scalar and its phantom counterpart is its contribution to the long-range force. The long-range mutual force between the two identical spacetimes is given by
\be
\lim_{r\rightarrow \infty} r^2\, \hbox{Force} = -M^2 - \varepsilon \Sigma^2\,,\label{longrange}
\ee
where $\varepsilon=1$ for a regular massless scalar and $\varepsilon=-1$ for a phantom massless scalar. Two identical Ellis wormholes or Ellis-Bronnikov wormholes are always repulsive, since the scalar hair is bigger than the mass, namely $\Sigma>M$. By contrast, two identical spacetime configurations carrying mass and a normal scalar hair necessarily always attract.

The no-hair theorem associated with a free massless scalar can be easily evaded by a variety of means, including but not limited to introducing an appropriate scalar potential, gauge fields, {\it etc}. (See a review \cite{Herdeiro:2015waa}.) One prominent procedure is the curvature-induced scalarization, where the massless scalar is non-minimally coupled to the topological Gauss-Bonnet term \cite{Doneva:2017bvd,Silva:2017uqg,Antoniou:2017acq}. (See also earlier string-inspired related works \cite{Kanti:1995vq}.) In addition to the Schwarzschild black hole, the theory admits new scalar-hairy black holes within a suitable mass range where the scalar hair parameter $\Sigma$ is a function of the mass. The procedure was extended to a massive scalar field a la Gauss-Bonnet extension of the Starobinsky gravity \cite{Liu:2020yqa}. In this paper, we consider the same procedure for the phantom massless scalar. New black holes beyond the Schwarzschild black hole must exist even for the phantom scalar. The reason is as follows. The black hole scalarization mechanism and hence the search for new black holes are based on the phenomenon that the linearized scalar develops instability in the Schwarzschild black hole background at a suitable critical mass $M_{\rm cr}$, where a new scalar hairy black hole emerges.  The equation of motion for the linearized scalar on the background of the Schwarzschild metric is independent of whether the scalar is standard or a phantom. Indeed, our numerical results indicate that the curvature-induced scalarization works equally well for the phantom scalar. We compare the properties of the black holes in Einstein-scalar-Gauss-Bonnet (ESGB) and Einstein-phantom-scalar-Gauss-Bonnet (EPSGB) gravities.

Black holes (or nonsingular solitons) involving phantom scalars were also constructed in literature, either with suitable scalar potential and/or with Maxwell fields that are non-minimally coupled to the scalar \cite{Gibbons:1996pd,Bronnikov:2005gm,Gao:2006iw,
Dzhunushaliev:2008bq,Jamil:2008bc,Clement:2009ai}. The curvature-induced scalarization mechanism has a particularly simple structure, without introducing new charges, at the price of not having an exact solution. Recently, a weak no-hair theorem conjecture was proposed for a black hole that involves scalar or other non-gauge fields. It states that the most general solution involving all allowed independent hairy parameters is necessarily not a black hole, which requires a fine-tuning of these hairy parameters \cite{Lu:2025eub}. For the ESGB or EPSGB theories, the scalar is real and the only non-gauge field and hence the most general spherically-symmetric and static solution contains two independent parameters $(M,\Sigma)$, which are called primary hair in literature. The weak no-hair theorem conjecture states that this general solution cannot be a black hole, and the black hole requires a fine-tuning of $\Sigma=\Sigma(M)$, which is called secondary scalar hair in literature. We compare this $\Sigma(M)$ function in ESGB and EPSGB gravities.

One important difference between ESGB and EPSGB is that black holes with negative mass can emerge in the latter theory. The fact that black holes carrying phantom hair can have negative mass was also observed in \cite{Lu:2015cqa,Lu:2015psa,Huang:2022urr}, where novel solutions carrying ghostlike massive spin-2 hair was constructed. Analogously, the massive spin-2 hair is secondary, depending on the mass of the black hole. From the point of view of black holes, the Gauss-Bonnet curvature induces a secondary scalar hair. On the other hand, this mechanism actually descalarizes the wormhole's independent primary scalar hair to the secondary one. In particular, we find that black holes with negative mass emerge in the EPSGB gravity. For these solutions, black hole scalarization is not a suitable description, since there is no Schwarzschild black hole with negative mass. However, there can be wormholes with negative mass and an independent scalar hair. Descalarization of the wormhole to become a black hole of the same mass is a more suitable description for this phenomenon.

The paper is organized as follows. In section 2, we introduce EPSGB gravity and study both the asymptotic and horizon structure. The general solutions contain two nontrivial parameters, the mass $M$ and scalar charge $\Sigma$. In section 3, we review the global structure of the general spherically-symmetric and static solutions of Einstein gravity minimally coupled to a massless phantom scalar. In addition to the usual Ellis-Bronnikov wormhole with $\Sigma>M$, we also study the global structure of solutions with $M\ge \Sigma$. We put some effort into this section since black holes with $M> \Sigma$, $M=\Sigma$ and $M<\Sigma$ can all rise in EPSGB gravity. In section 4, we present the numerical black hole solutions and summarize their properties. We conclude the paper in section 5.  Since our black hole construction is numerical, the results are illustrated by a large number of graphs, which we put all in the appendix.

\section{Einstein-Phantom-Scalar-Gauss-Bonnet gravity}

\subsection{The theory}

We begin with Einstein gravity minimally coupled to a massive free scalar field. We then consider an additional non-minimal coupling between the scalar field and the Gauss-Bonnet combination.
The Lagrangian is
\begin{equation}
	\label{eq1001}
	I=\int \text{d}^4 x \sqrt{-g}\Big[R+\varepsilon \Big(-\frac12(\partial\phi)^2-\frac12\mu^2\phi^2+U(\phi)
(R^2-4R_{\mu\nu}R^{\mu\nu}+R_{\mu\nu\rho\sigma}R^{\mu\nu\rho\sigma})\Big)\Big],
\end{equation}
where $\varepsilon=\pm 1$. For $\varepsilon=1$, the scalar field has the standard kinetic term and the theory is the standard Einstein-Scalar-Gauss-Bonnet gravity (ESGB), with $U(\phi)$ a generic coupling function between the scalar and the Gauss-Bonnet term. In this paper, we shall focus instead on the case with $\varepsilon=-1$, where the scalar has the phantom-like kinetic term. The theory can be described as Einstein-Phantom-Scalar-Gauss-Bonnet gravity (EPSGB).

The theory \eqref{eq1001} has two fundamental fields, the metric and the scalar. Their variation principle leads to two covariant equations of motion
\begin{equation}
	\label{eq1002}
	\begin{aligned}
		0=&\: R_{\mu\nu}-\frac12Rg_{\mu\nu} + \varepsilon\Big(-2R\nabla_\mu\nabla_\nu U(\phi)-4\square U(\phi)(R_{\mu\nu}-\frac12Rg_{\mu\nu})\\
		&+4R_{\mu\alpha}\nabla^\alpha\nabla_\nu U(\phi)+4R_{\nu\alpha}\nabla^\alpha\nabla_\mu U(\phi)-4g_{\mu\nu}R^{\alpha\beta}\nabla_\alpha\nabla_\beta U(\phi)\\
		&+4R^\beta{}_{\mu\alpha\nu}\nabla^\alpha\nabla_\beta U(\phi)-\frac12\partial_\mu\phi\partial_\nu\phi+
\frac14(\partial\phi)^2g_{\mu\nu}+\frac14\mu^2\phi^2g_{\mu\nu}\Big)\:,\\
		0=&\: \square\phi-\mu^2\phi-\frac{\delta U(\phi)}{\delta\phi}\Big(R^2-4R_{\mu\nu}R^{\mu\nu}+R_{\mu\nu\rho\sigma}R^{\mu\nu\rho\sigma}\Big)\:.
	\end{aligned}
\end{equation}
For simplicity, we consider only the massless scalar field $(\mu=0)$ throughout this work. While black hole and wormhole solutions with $\varepsilon = 1$ have been widely investigated, {\it e.g.}, \cite{Kanti:1995vq,Doneva:2017bvd,Silva:2017uqg,Antoniou:2017acq}, we focus on the case of a phantom scalar field with $\varepsilon = -1$. Furthermore, we shall adopt the scalar coupling function $U$ as \cite{Doneva:2017bvd}
\begin{equation}
	\label{eq2001}
	U(\phi)=\frac{\lambda^2}{12}(1-e^{-6\phi^2}),
\end{equation}
where $\lambda$ is a coupling constant. It satisfies the condition
\begin{equation}
	\label{eq2002}
	\frac{\delta U(\phi)}{\delta\phi}\Big|_{\phi=0}=0\:,\qquad
\frac{\delta^2U(\phi)}{\delta^2\phi}\Big|_{\phi=0}=\lambda^2>0\:.
\end{equation}
Thus the Schwarzschild black hole remains an exact solution of the theory. Furthermore, one can construct new black hole solutions carrying scalar hair. This is called curvature induced scalarization. However, it should be emphasized that the black hole solution does not have a free scalar hair parameter, which is instead a fixed function of the mass.

\subsection{Spherically-symmetric and static ansatz}

We consider the spherically-symmetric and static spacetimes, focusing on the construction of new black holes. The most general ansatz can be written as
\begin{equation}
	\label{eq20021}
	\text{d}s^2=-h(r)\text{d} t^2+\frac{\text{d} r^2}{f(r)}+r^2(\text{d}\theta^2+\sin\theta^2\text{d}\varphi^2)\:,\qquad\phi=\phi(r)\:.
\end{equation}
where $h(r)$, $f(r)$, and $\phi(r)$ are three functions to be determined. As in the Schwarzschild metric, the radius of the foliating 2-sphere is treated as the radial coordinate. We call this $r$ as the Schwarzschild-like radial coordinate. Substituting the ansatz \eqref{eq20021} into the field equations \eqref{eq1002}, we obtain three coupled nonlinear differential equations for the three functions:
\begin{equation}
	\label{eq20022}
	\begin{aligned}
		&-4 \lambda ^2 h(r) \phi (r) f'(r) h'(r)+12 \lambda ^2 f(r) h(r) \phi (r) f'(r) h'(r)+r^2 h(r)^2 e^{6 \phi (r)^2} f'(r) \phi '(r)\\
		&+8 \lambda ^2 f(r)^2 h(r) \phi (r) h''(r)-8 \lambda ^2 f(r) h(r) \phi (r) h''(r)+r^2 f(r) h(r) e^{6 \phi (r)^2} h'(r) \phi '(r)\\
		&-4 \lambda ^2 f(r)^2 \phi (r) h'(r)^2+4 \lambda ^2 f(r) \phi (r) h'(r)^2+2 r^2 f(r) h(r)^2 e^{6 \phi (r)^2} \phi ''(r)\\
		&+4 r f(r) h(r)^2 e^{6 \phi (r)^2} \phi '(r)=0,\\
		&-8 \varepsilon  \lambda ^2 \phi (r) f'(r) \phi '(r)+24 \varepsilon  \lambda ^2 f(r) \phi (r) f'(r) \phi '(r)-4 r e^{6 \phi (r)^2} f'(r)\\
		&-\varepsilon  r^2 f(r) e^{6 \phi (r)^2} \phi '(r)^2+16 \varepsilon  \lambda ^2 f(r)^2 \phi (r) \phi ''(r)-16 \varepsilon  \lambda ^2 f(r) \phi (r) \phi ''(r)\\
		&+16 \varepsilon  \lambda ^2 f(r)^2 \phi '(r)^2-192 \kappa  \lambda ^2 f(r)^2 \phi (r)^2 \phi '(r)^2+192 \varepsilon  \lambda ^2 f(r) \phi (r)^2 \phi '(r)^2\\
		&-16 \varepsilon  \lambda ^2 f(r) \phi '(r)^2-4 f(r) e^{6 \phi (r)^2}+4 e^{6 \phi (r)^2}=0,\\
		&-12 \varepsilon  \lambda ^2 f(r) h(r) \phi (r) f'(r) h'(r) \phi '(r)+r h(r) e^{6 \phi (r)^2} f'(r) h'(r)+2 h(r)^2 e^{6 \phi (r)^2} f'(r)\\
		&-8 \varepsilon  \lambda ^2 f(r)^2 h(r) \phi (r) h''(r) \phi '(r)+2 r f(r) h(r) e^{6 \phi (r)^2} h''(r)\\
		&-8 \varepsilon  \lambda ^2 f(r)^2 h(r) \phi (r) h'(r) \phi ''(r)+96 \varepsilon  \lambda ^2 f(r)^2 h(r) \phi (r)^2 h'(r) \phi '(r)^2\\
		&-8 \varepsilon  \lambda ^2 f(r)^2 h(r) h'(r) \phi '(r)^2+4 \varepsilon  \lambda ^2 f(r)^2 \phi (r) h'(r)^2 \phi '(r)-r f(r) e^{6 \phi (r)^2} h'(r)^2\\
		&+2 f(r) h(r) e^{6 \phi (r)^2} h'(r)+\varepsilon  r f(r) h(r)^2 e^{6 \phi (r)^2} \phi '(r)^2=0,
	\end{aligned}
\end{equation}

\subsection{Asymptotic falloffs}

In the large-$r$ asymptotic regions, the general expansions of the falloffs of the metric and scalar functions are given by
\begin{equation}
	\label{eq2003}
	\begin{aligned}
		&h\:=\:h_0+\frac{h_1}r+\frac{h_2}{r^2}+\frac{h_3}{r^3}+\ldots,\\
		&f\:=\:f_0+\frac{f_1}r+\frac{f_2}{r^2}+\frac{f_3}{r^3}+\ldots,\\
		&\phi\:=\:\phi_0+\frac{\phi_1}r+\frac{\phi_2}{r^2}+\frac{\phi_3}{r^3}+\ldots,
	\end{aligned}
\end{equation}
The constant coefficients $h_i$, $f_i$, and $\phi_i$ can be solved order by order in the large-$r$ expansion, with the low-lying examples given by
\begin{equation}
	\label{eq2004}
	\begin{aligned}
	&f_{0}=1,\qquad h_{1}=h_{0}f_{1},\qquad h_{2}=0,\quad f_2=\frac{\varepsilon \phi_1^2}{4},\quad\phi_2=-\frac{f_1\phi_1}{2},\\
	&h_{3}=-\frac{\varepsilon}{24}f_1h_0\phi_1^2,\qquad f_3=-\frac{\varepsilon f_1\phi_1^2}{8},\quad\phi_3=\frac{8f_1^2\phi_1-\varepsilon \phi_1^3}{24},\\
	&h_{4}=\frac{\varepsilon}{24}f_1h_0\phi_1(f_1\phi_1-48e^{-6\phi_0}\phi_0\lambda^2),\quad f_4=\frac{\varepsilon}{12}f_1\phi_1(f_1\phi_1-48e^{-6\phi_0}\phi_0\lambda^2),\\
	&\phi_{4}=-\frac{f_1^3\phi_1}{4}+\frac{\varepsilon f_1\phi_1^3}{12}+e^{-6\phi_0^2}f_1^2\phi_0\lambda^2.
	\end{aligned}
\end{equation}
Note that the equations of motion fix $f_0=1$. We can scale the time coordinate appropriately so that $h_0=1$. Thus, we see that the general asymptotic Minkowski spacetimes are specified by three parameters, $(f_1,\phi_0,\phi_1)$. The parameter $f_1$, associated with mass, originates from the metric functions, whilst the parameters $(\phi_0,\phi_1)$ describe the scalar hair. Interestingly, the $\phi_0$ parameter enters the metric only at the 4'th order of the falloffs. However, for constructing black holes, we find that $\phi_0$ vanishes. The mass and scalar hair are given by
\be
M=-\ft12 f_1\,,\qquad \Sigma = \ft12 \phi_1\,.\label{masshair}
\ee
The long-range force between two identical black holes is given by \eqref{longrange}.

\subsection{Near horizon geometry}

We have seen that the asymptotically-flat solutions are parameterized by three parameters. The general solution does not describe a black hole. A numerical analysis indicates that for given $f_1$, we need to fine-tune the scalar hair parameters $(\phi_0,\phi_1)$ carefully, so that the asymptotic flat geometry can be integrated into an interior surface at $r=r_0$, where $f(r_0)=0=h(r_0)$, with $\phi(r_0)$ finite. The equations of motion \eqref{eq20022} become singular at $r=r_0$ under the Schwarzschild-like radial coordinate. In order to examine further the spacetime region $r <r_0$, it is necessary for us to study the near-horizon geometry. We assume that the three functions near $r=r_0$ are analytic, with the Taylor expansions
\bea
h &=& h_1^+ (r-r_+) + h_2^+ (r-r_+)^2 + h_3^+ (r-r_+)^3 +\cdots\,,\nn\\
f&=&f_1^+ (r-r_+) + f_2^+ (r-r_+)^2 + f_3^+ (r-r_+)^3+ \cdots\,,\nn\\
\phi&=&\phi_0^+ + \phi_1^+ (r-r_+) + \phi_2^+ (r-r_+)^2 + \phi_3^+ (r-r_+)^3 + \cdots\,.
\eea
Note that the parameter $h_1^+$ is locally trivial and can be set to 1 by an appropriate time scaling. Substituting these into the equations of motion, we can solve the coefficients order by order. The lowest order equations require the constraint
\be
r_+^3 \phi_1^+ e^{2(\phi_0^+)^2}+ 2\lambda^2 \phi_0^+ \Big(6 + \varepsilon r_+^2 (\phi_1^+)^2\Big)=0\,.
\ee
The remaining coefficients can be solved order by order with increasing complexity. The $f_1=1/r_+ + \varepsilon r_+ (\phi_1^+)^2/6$ is relatively simple, but the higher-order coefficients become very complicated and we shall not present them here. For numerical accuracies, we perform the power expansion up to and including the ninth order. The upshot is that the near-horizon geometry is specified by two independent parameters $(r_+,\phi_0^+)$, or alternatively $(r_+, \phi_1^+)$, i.e.~the horizon radius and a scalar hair parameter. This seems to suggest a possibility that we can construct a black hole with a free scalar parameter, in addition to the mass. This turns out not to be the case by numerical analysis which shows that for given $r_0$, we have to fine-tune the scalar hair parameter carefully in order to integrate the near-horizon geometry to the asymptotically-flat region.

\section{Einstein gravity coupled to a free phantom scalar}

In this section, we review spherically-symmetric and static solutions in Einstein gravity coupled to a free scalar. The Lagrangian is given by \eqref{eq1001} with $\lambda=0$. This is the ideal case to apply the no-hair theorem. The free scalar field cannot be turned on in a black hole solution and hence the only black hole is the Schwarzschild black hole. However, since the theory is sufficiently simple, one can construct exact solutions of the general spherically-symmetric and static geometry. For the normal free scalar with $\varepsilon=1$, the solution was given in \cite{Janis:1968zz}. The solution is asymptotically flat containing three integration constants. In our language, they are $(f_1,\phi_0,\phi_1)$, the asymptotic falloff parameters discussed in the previous section. For the free scalar, the parameter $\phi_0$ is trivial, reflecting the constant shift symmetry of a free scalar. The general solution contains a naked singularity, and reduces to the Schwarzschild black hole when we turn off the scalar field. The curvature-induced scalar hairy black holes were constructed in \cite{Doneva:2017bvd,Silva:2017uqg,Antoniou:2017acq}, where scalar hair parameter becomes a specific function of mass.

In this paper, we shall focus on the second case, where the scalar is a phantom with $\varepsilon=-1$. The general solution of Einstein gravity coupled to such a free phantom scalar describes a smooth (Ellis-Bronnikov) wormhole connecting two asymptotic Minkowski spacetimes, for an appropriate choice of the integration constants. The original solution contains two integration constants $(m,q>0)$ and the wormhole spacetime requires $|m|<q$. (The trivial $\phi_0$ constant can be set to zero.) This implies the mass and scalar hair relation is $\Sigma>M$. However, we find that black holes with $M\ge \Sigma$ can also emerge in EPSGB gravity. It is thus of interest to study also the global structure of the solutions with $M\ge \Sigma$ in Einstein-phantom-scalar (EPS) gravity.

\subsection{$-q<m<q$}

The Ellis-Bronnikov wormhole metric is given by \cite{Bronnikov:1973fh}
\be
ds^2 = -e^{\fft{m}{q}\bar\phi} dt^2+ e^{-\fft{m}{q}
\bar \phi} d\tilde r^2+R^2(\tilde r)d\Omega_{2}^2\,,\qquad R^2=e^{-\fft{m}{q} \bar\phi}\,(\tilde r^2+q^2-m^2)\,.\label{ebmetric}
\ee
The free massless phantom scalar is
\be
\phi=\phi_0+\bar\phi\,,\qquad
\bar\phi = \fft{2q}{\sqrt{q^2-m^2}} \arctan\Big(\fft{\tilde r}{\sqrt{q^2-m^2}}\Big)\,.
\ee
Note that when a scalar field is involved, the Schwarzschild-like radius is not the best coordinate to describe an exact solution. The solution contains one trivial parameter $\phi_0$ and two nontrivial parameters, $(m,q)$, both parameters can be both positive and negative. We can choose without loss of generality that $q$ is positive, but the parameter $m$ can be either sign, satisfying the constraint $|m|< q$.  The wormhole throat is located at $\tilde r= m$, corresponding to the wormhole radius $a$
\be
a^2 = R_{\rm min}^2 = q^2 e^{-\frac{2 m \tan ^{-1}\left(\frac{m}{\sqrt{q^2-m^2}}\right)}{\sqrt{q^2-m^2}}}\,.
\ee

We can now write the Schwarzschild-like coordinate $R=r$, then the metric is cast into the following form
\be
ds^2 = - h dt^2 + \fft{dr^2}{f} + r^2 d\Omega^2\,,\qquad \phi=\phi(r)\,.\label{schcoord}
\ee
For large $r$, we have
\bea
h &=& e^{\frac{\pi  m}{\sqrt{q^2-m^2}}} \Big(1-\frac{2m e^{-\frac{\pi  m}{2 \sqrt{q^2-m^2}}}}{r}-\frac{m q^2 e^{-\frac{3 \pi  m}{2 \sqrt{q^2-m^2}}}}{3 r^3}-\frac{2m^2 q^2 e^{-\frac{2 \pi  m}{\sqrt{q^2-m^2}}}}{3 r^4}+O\left(\left(\ft{1}{r}\right)^5\right)\Big)\,,\nn\\
f &=& 1-\frac{2m e^{-\frac{\pi  m}{2 \sqrt{q^2-m^2}}}}{r}-\frac{q^2 e^{-\frac{\pi  m}{\sqrt{q^2-m^2}}}}{r^2}-\frac{m q^2 e^{-\frac{3 \pi  m}{2 \sqrt{q^2-m^2}}}}{r^3}-\frac{m^2 q^2 e^{-\frac{2 \pi  m}{\sqrt{q^2-m^2}}}}{3 r^4}+O\left(\left(\ft{1}{r}\right)^5\right),\nn\\
\phi &=& \left(\pi  q \sqrt{\frac{1}{q^2-m^2}}+\phi_0\right)-\frac{2q e^{-\frac{\pi  m}{2 \sqrt{q^2-m^2}}}}{r}-\frac{m q e^{-\frac{\pi  m}{\sqrt{q^2-m^2}}}}{r^2}-\frac{q e^{-\frac{3 \pi  m}{2 \sqrt{q^2-m^2}}} \left(8 m^2+q^2\right)}{3 r^3}\nn\\
&&-\frac{4 m q e^{-\frac{2 \pi  m}{\sqrt{q^2-m^2}}} \left(3 m^2+q^2\right)}{3 r^4}+O\left(\left(\ft{1}{r}\right)^5\right).\label{falloff1}
\eea
In terms of the Schwarzschild-like radial coordinate $r$, the wormhole throat is located at $r=a$, with
\be
h(a)=\fft{q^2}{a^2} e^{-\frac{\pi  m}{\sqrt{q^2-m^2}}}\,,\qquad f(a)=0\,.
\ee
Note that here, we have rescaled the time coordinate so that $h(\infty)=1$.

The global structure of the wormhole with general $(m,q)$ was analyzed in \cite{Huang:2020qmn}. In particular, it was seen that there is no global scaling so that the speed of light can be set to be the same for both $r\rightarrow \pm \infty$ asymptotic regions, except for $m=0$, corresponding to the Ellis wormhole. This implies that the wormhole cannot be used to tunnel between two regions of the same Minkowski spacetime. Furthermore, the mass can be both positive and negative; if the wormhole mass measured in one asymptotic region is positive, it is negative measured in the other region. The emergence of negative mass in wormholes were also noted in \cite{Hao:2023kvf}. It follows from \eqref{longrange} that the long-range force between two identical such wormholes is always repulsive.

\subsection{$m>q$}

In this case, the metric of the solution takes the same form as \eqref{ebmetric}, but the scalar field now becomes
\be
\bar\phi = -\fft{2q}{\sqrt{m^2-q^2}} {\rm arctanh}(\fft{\sqrt{m^2-q^2}}{\tilde r})\,,
\ee
The asymptotic flat Minkowski region is at $\tilde r\rightarrow \infty$. There is a curvature singularity at
\be
\tilde r_*=\sqrt{m^2-q^2}\,.
\ee
To see the singularity structure, it is useful to define a ``co-moving'' length coordinate $\rho$, given by
\be
\rho\sim (\tilde r-\tilde r_*)^{\fft{m + 4 \sqrt{m^2-q^2}}{4\sqrt{m^2-q^2}}} \rightarrow 0\,.
\ee
The metric near the singularity becomes
\be
ds^2\sim -\rho^{\fft{4m}{m + \sqrt{m^2-q^2}}} dt^2 + d\rho^2 +
\rho^{-\fft{4(m-\sqrt{m^2-q^2})}{m + 4\sqrt{m^2-q^2}}} d\Omega_2^2\,.
\ee
Thus, the curvature power-law singularity is characterized by the blowing up of the $S^2$ radius at some finite co-moving length. In particular, since we have
\be
\fft{4m}{m + \sqrt{m^2-q^2}}> 2\,,
\ee
there is an infinite red shift at the singularity. We may call this a null singularity.

The radius of the foliating $S^2$ blows up at both the singularity $\tilde r^*$ and asymptotic infinity. It follows that in between $(\tilde r_*, \infty)$, there must be a wormhole throat, located at $\tilde r=m>r^*$, and the corresponding wormhole radius is
\be
a=R_{\rm min}=q e^{\frac{m \tanh ^{-1}\left(\frac{\sqrt{m^2-q^2}}{m}\right)}{\sqrt{m^2-q^2}}}\,.
\ee
Thus, we see that when $m\ge q$, the solution describes a wormhole connecting Minkowski spacetime $\tilde r\rightarrow \infty$ to a curvature singularity at $\tilde r=\tilde r_*$.  In terms of the Schwarzschild-like coordinate, the large $r\sim \tilde r$ behavior is
\bea
h &=& 1-\frac{2 m}{r}-\frac{m q^2}{3 r^3}-\frac{2 \left(m^2 q^2\right)}{3 r^4}+O\left(\left(\frac{1}{r}\right)^5\right)\,,\nn\\
f &=& 1-\frac{2 m}{r}-\frac{q^2}{r^2}-\frac{m q^2}{r^3}-\frac{4 \left(m^2 q^2\right)}{3 r^4}+O\left(\left(\frac{1}{r}\right)^5\right)\,,\nn\\
\phi &=& -\frac{2 q}{r}-\frac{2 (m q)}{r^2}-\frac{q \left(8 m^2+q^2\right)}{3 r^3}-\frac{4 \left(m q \left(3 m^2+q^2\right)\right)}{3 r^4}+O\left(\left(\frac{1}{r}\right)^5\right)\,.
\label{falloff2}
\eea
In this Schwarzschild-type coordinate system, the wormhole throat is located at $r=a$, for which we have
\be
h(a)=\fft{q^2}{a^2}\,,\qquad f(a)=0\,.
\ee
The long-range force between two identical configurations is attractive, since we have $M>\Sigma$.

\subsection{$m=q$}

Note that there is a smooth limit to $m=q$, in which case, the solution is given by
\be
\bar \phi=-\fft{2q}{\tilde r}\,.
\ee
The radius of the foliating $S^2$ is $R=\tilde r e^{q/\tilde r}$. As the radial coordinate runs from 0 to $\infty$, $R$ decreases from infinity to a minimum value at $r=q$ with $R_{\rm min} = q \exp(1)$, and then increases to infinity again. The $\tilde r=0$ is not a singularity, but locally flat, since all curvature tensors vanish. However, this is not the asymptotic Minkowski region either, with $g_{tt}\rightarrow 0$. Thus, the spacetime geometry describes a wormhole connecting Minkowski spacetime ($\tilde r\rightarrow \infty$) to a locally flat region $\tilde r\rightarrow 0$ with infinite red shift. The wormhole throat is at $\tilde r=q$, with wormhole radius $R_{\rm min}$. This type of geometry was referred to as a dark wormhole, as the other side of the world is non-singular but dark owing to infinite red shift \cite{Geng:2015kvs}. The long-range force between such two identical configurations vanishes. Finally, for $m<-q$, the asymptotically-flat solution simply has a naked singularity.

It is worth commenting that the asymptotic falloff expansions \eqref{falloff1} for $|m|<q$ and \eqref{falloff2} for $m>q$ are formally the same after some appropriate reparameterization. In the Schwarzschild-like radial coordinate system, both cases give rise to the wormhole throat $r=a$, where $h(a)>0=f(a)$. Thus in this coordinate system, if we have only numerical solution, we cannot deduce what is the qualitative geometric structure in the other side of the wormhole. The explicit example tells us that the other side could be asymptotic to a different Minkowski spacetime for $|m|<q$, or a singularity of infinite $S^2$ radius for $m> q$. We can illustrate this explicitly using the $m=q$ (dark wormhole) example, where the solution can be expressed exactly in the Schwarzschild-like radial coordinate:
\bea
\hbox{this side}:\qquad ds^2 &=& - e^{2 W_0\left(-\frac{q}{r}\right)} dt^2 + \fft{dr^2}{\left(W_0\left(-\frac{q}{r}\right)+1\right)^2} + r^2 d\Omega_2^2\,,\quad
\phi = 2 W_0\left(-\frac{q}{r}\right),\\
\hbox{other side}:\qquad ds^2 &=& - e^{2 W_{-1}\left(-\frac{q}{r}\right)} dt^2 + \fft{dr^2}{\left(W_{-1}\left(-\frac{q}{r}\right)+1\right)^2} + r^2 d\Omega_2^2\,,\quad
\phi = 2 W_{-1}\left(-\frac{q}{r}\right),\nn
\eea
where $W_0$ and $W_{-1}$ are the Lambert-$W$ product-log functions. The two metrics describe the two sides of the wormhole geometries in the Schwarzschild-like radial coordinate, and they join at the wormhole throat at $r=q \exp(1)$. The naive curvature singularity at $r=0$ is completely outside of the dark wormhole spacetime region.

For simplicity, we shall refer to all these three classes of solutions as wormholes. These solutions are specified by two independent continuous parameters $(M,\Sigma)$. Next, we show numerically that all these three classes of wormhole solutions can be descalarized to become a black hole with $\Sigma=\Sigma(M)$ for a suitable mass range.

\section{Numerical results}

To obtain black hole solutions numerically, we implement the shooting method by integrating the equations inward from asymptotic infinity until both metric functions $h(r)$ and $f(r)$ simultaneously approach zero on the horizon. We fix our parameter configuration with these principles: we set $\varepsilon=-1$ to ensure a negative kinetic term for the phantom scalar field. However, we also reconstruct the $\varepsilon=1$ results of \cite{Doneva:2017bvd} so as to confirm our numerical method, as well as to compare the $\varepsilon=\pm 1$ black holes. The coupling constant  $\lambda$ is only the dimensionful parameter of the theory and we can use it to construct dimensionless quantities. Thus, without loss of generality, we fix $\lambda=2$ for computational purposes. The Minkowski vacuum is specified by $\phi_{0}=0$ \cite{Doneva:2017bvd}. Numerically, we find that black hole solutions only exist when $\phi_{0}=0$. The time scaling implies that we can fix $h_{0}=1$, so that the speed of light at infinity is set to 1. These restrictions leave only two nontrivial parameters $f_{1}$ and $\phi_{1}$ to be determined, giving rise to two independent parameters, mass and scalar charge, defined by \eqref{masshair}.  For a real scalar, the weak no-hair theorem conjecture implies that $\Sigma$ is not an independent parameter, but secondary, depending on the mass, namely $\Sigma=\Sigma(M)$. Our numerical results confirm this conjecture. Specifically, when using the shooting method, we first select a number for $f_{1}$ and then carefully fine-tune the parameter $\phi_{1}$ until both $h(r)$ and $f(r)$  simultaneously vanish at some finite $r_{0}>0$.

\subsection{Summary}

\noindent{\bf Black hole profiles:} Our numerical analysis reveals the existence of black hole solutions with positive, zero, and negative masses. Fig.~\ref{Fig20021} in Appendix displays the exterior (outside of the horizon) radial profiles of the metric functions $h(r)$, $f(r)$, and the scalar field $\phi(r)$ for four representative mass values $M/\lambda=-0.01, 0, 0.01$, and $0.2$. We see that for negative or smaller mass, the scalar contribution is more prominent and the black holes resemble less of the Schwarzschild black hole. As mass increases, the scalar influence wanes and the black hole becomes more like the Schwarzschild black hole. As we shall see presently, there is a maximum mass or critical mass $M_{\rm cr}$ beyond which only the Schwarzschild black hole exists.

We can further extend our numerical integration inward through the event horizon and obtain the black hole interior structure. We find that both metric functions $h(r)$ and $f(r)$ both diverge to $-\infty$ as $r \to 0$, indicating the presence of a space-like curvature singularity analogous to the Schwarzschild case. This singular behavior is demonstrated in Fig.~\ref{Fig200211}, which shows the interior profiles of $h(r)$, $f(r)$, and $\phi(r)$ for the four different mass parameters. Despite having a phantom scalar, the scalar hairy black holes are singular, without the black bounce structure of \cite{Simpson:2018tsi}.

\noindent{\bf General profiles:} The scalar hair parameter has to be carefully fine-tuned to construct a black hole. It is of interest to examine the solution profiles for general parameters. Here we present the negative mass solutions with fixed parameter $f_1 = 0.02$.  Fig.~\ref{Fig20038} shows three characteristic cases with $\phi_1 = 0.418$, $0.428$, and $0.438$, corresponding respectively to:
\begin{itemize}
	\item ``Almost regular'' horizonless spacetime
	\item Black hole solution with an event horizon
	\item Wormhole solution
\end{itemize}
Note that when $\phi_1<0.428$, all functions $(h,f,\phi)$ are finite as $r$ runs from 0 to $\infty$. However, the solution is still singular since regularity requires that $f(0)=1$. We call such configuration almost regular horizonless spacetime since the power-law curvature singularity is severely subdued.

For positive mass, the general profiles are quite different. Figure~\ref{Fig20039} shows four representative solutions with $\phi_1$ values of $0.524$, $0.794$, $0.824$, and $0.994$, demonstrating:
\begin{itemize}
	\item Wormhole solution (for $\phi_1 = 0.524$ and $0.994$)
	\item Singular spacetime with zero $h(r)$ but finite and nonzero $f(r)$ ($\phi_1 = 0.794$)
	\item Black hole solution ($\phi_1 = 0.824$)
\end{itemize}

\noindent{\bf Black hole mass spectrum:} By scanning the parameter $ f_1 $, we numerically construct all scalar hairy black hole solutions. We identify a critical mass $ M_{\rm cr}=M_{\text{max}} \approx 1.173 $ with a corresponding horizon radius $ r_0 \approx 2.345 $, see Fig.~\ref{Fig20022}.  Below this threshold, larger-mass solutions approach the Schwarzschild metric asymptotically, while no scalarized solutions exist for $ M \geq M_{\text{max}} $. Beyond this mass, the only spherically-symmetric and static black hole is the Schwarzschild black hole.
In other words, $M_{\text{max}}$ is the critical scalarization mass $M_{\rm cr}$ mentioned in the introduction.

Notably, switching $ \varepsilon=-1 $, representing EPSGB gravity, to $\varepsilon=1 $, representing ESGB, preserves the critical mass $ M_{\rm cr}=M_{\text{max}} $, but the mass-horizon radius relations differ:
\begin{itemize}
	\item \textbf{EPSGB}: Scalarized solutions always have $ M < M_{\text{Schwarzschild}} $ and admit negative masses ($ M_{\text{min}} \approx -0.01 $).
	\item \textbf{ESGB}: Scalarized solutions satisfy $ M > M_{\text{Schwarzschild}} $ with $ M \geq 0 $.
\end{itemize}
This stark contrast highlights the role of $ \varepsilon $ in determining the allowed mass range and the dominance of scalarized solutions over Schwarzschild in different regimes. The EPSGB theory's negative-mass solutions, absent in ESGB, further distinguish the two theories.

\noindent{\bf Long-range force:} One important aspect of black holes is the long-range force between two identical black holes, specified by the mass and scalar hair parameter, given in \eqref{longrange}. For the ESGB theory, the long-range force is always attractive, but situation is different in the EPSGB theory. Due to the $\mathbb{Z}_2$ symmetry ($\phi \leftrightarrow -\phi$) of the scalar field, $\phi_1$ exhibits sign ambiguity, reflecting the degenerate nature of solutions. We define $\Sigma = \phi_1/2$. Fig.~\ref{Fig20032} shows the relationship between $\Sigma$ and the mass $M$ for both EPSGB ($\varepsilon=-1$) and ESGB ($\varepsilon=1$) theories. For the ESGB scalarized black holes, the $\Sigma$-$M$ curve forms a closed loop, indicating a finite range of allowed scalar charges for $0 \leq M \leq M_{\text{max}}$. In contrast, for the EPSGB scalarized black holes, the curve fails to close at low masses ($M \to M_{\text{min}}$), extending to negative masses (see Fig.~\ref{Fig20032}). The non-closure suggests a qualitative difference between the solutions.  Both curves share the same maximum mass, consistent with the mass-radius results in Figure~\ref{Fig20022}. This is expected since the maximum mass is precisely the critical mass, discussed in the introduction. The structural difference in the $\Sigma$/$M$ relation underscores how the sign of $\varepsilon$ influences the scalar-hairy behavior, particularly near the low mass. Since all the three cases, namely $\Sigma<M$, $\Sigma=M$ and $\sigma >M$, can arise in black holes, the long-range force between two identical black holes in EPSGB gravity can be attractive, zero or repulsive respectively.

\noindent{\bf Black hole thermodynamics:} We now examine the thermodynamic properties of these scalarized black holes. The Hawking temperature can be derived from the surface gravity on the event horizon $r_0$. The entropy can be calculated using Wald entropy formula \cite{Wald:1993nt,Iyer:1994ys}. They are given by
\begin{equation}\label{tempentropy}
T = \frac{\sqrt{f'(r_0) h'(r_0)}}{4\pi}\,,\qquad S=\pi r_{0}^2 + 4 \pi \varepsilon U(\phi_{h})\,,
\end{equation}
where $\phi_{h}$ is the value of the scalar field $\phi(r)$ on the event horizon of the black hole. (Note that we also use $r_0$ to denote the horizon radius.)

By evaluating our numerical solutions at the event horizon, we extract the temperature $T$ and entropy $S$ for each black hole configuration. Fig.~\ref{Fig20033} shows both the temperature and entropy as functions of horizon radius $r_0$, demonstrating that the temperature and entropy of scalarized black holes coincide with the Schwarzschild values when the horizon radius reaches its maximum $r_0 \approx 2.345$.

Fig.~\ref{Fig200362} presents the further thermodynamic properties. The left panel illustrates the entropy-mass ($S$-$M$) relation, which shows that the entropy of EPSGB (ESGB) hairy black holes is always smaller (larger) than that of Schwarzschild black holes under the same mass. Furthermore, for EPSGB case, when the mass is approximately less than $M \approx 0.242$, the entropy becomes negative. In particular, at the minimum value of  $M \approx -0.10$, the minimum value of entropy is $S \approx -1.05$. Such negative entropy can be uplifted to a positive value with appropriate Gauss-Bonnet topological term that can affect the entropy, but not the equations of motion and hence the black hole solution.  We find that the numerical data can be well-described by the fitted relation:
\begin{equation}
	\label{eq:S_fit}
	S(M) = -1.04037 + 1.00952 M + 14.1698 M^2 - 2.71831  M^3 +\cdots\, ,
\end{equation}
which predicts an entropy minimum at $M \approx -0.035$ with $S_{\text{min}}^{\text{fit}} \approx -1.058$, in close agreement with the direct numerical results.  In the right panel of Fig.~\ref{Fig200362}, we present the relation of the free energy $F = M - TS$ as a function of temperature $T$. The results demonstrate that the EPSGB hairy black holes consistently exhibit higher free energy than Schwarzschild black holes at the same temperature. When the horizon radius reaches its maximum value (corresponding to the minimum temperature $T_{\text{min}}$), the free energy of the EPSGB hairy black holes coincides with that of Schwarzschild black holes. At higher temperatures, where the horizon radius decreases, the free energy of the hairy black holes increases monotonically with temperature - a behavior that contrasts sharply with the temperature dependence observed for Schwarzschild black holes.

Finally, we verify that the EPSGB scalarized black holes satisfy the first law of thermodynamics: $\text{d} M = T \text{d} S$. In Fig.~\ref{Fig20037}, the red curve represents $\text{d} M / \text{d} S$ obtained by differentiating the interpolating function $M(S)$, while the blue data points correspond to the temperature $T$ obtained from \eqref{tempentropy}. The excellent agreement between these quantities confirms the validity of the first law $\text{d} M = T \text{d} S$. Notably, the first law relation excludes a scalar term $\Psi d\phi$, because the EPSGB black holes lack an independent scalar-hairy parameter.

\section{Conclusions}

We investigated the most general spherically-symmetric and static solutions in Einstein gravity minimally coupled to a massless phantom scalar. The general solution contains two nontrivial independent parameters, the mass $M$ and scalar charge $\Sigma$. The well-known Ellis-Bronnikov  wormhole connecting two asymptotically-flat spacetimes corresponds to $\Sigma>M$, such that two identical such spacetimes experience repulsive force at the long range. We also studied the global structures when $M\ge \Sigma$, with the long-range force zero or attractive. There is also a wormhole throat that connects a flat spacetime to a singularity. When $M=\Sigma$, the singularity is null. For simplicity, we called all these spacetime configurations as wormholes.

One of the motivations of this work was based on the curiosity as what is the implication to black holes for the exotic matter that is responsible for the traversable wormholes. We adopted the curvature-induced scalarization mechanism for constructing the phantom scalar-hairy black holes. As explained in the introduction, this construction is guaranteed to succeed since it is based on the fact that new hairy black holes emerge at the critical Schwarzschild mass $M_{\rm cr}$ where the linearized scalar mode develops an instability. This phenomenon is independent of whether the scalar is regular or phantomlike. It turns out that the critical mass is the maximum mass $M_{\rm max}$ for the scalar hairy black hole for both regular and phantomlike scalars. Away from the critical mass, the scalar hairy black holes become very different for the EPSGB and ESGB gravities, and we made a comprehensive comparison.

From the Schwarzschild black hole point of view, the mechanism is a scalarization, but from the wormhole point of view, it is a descalarization procedure since the independent primary scalar hair parameter $\Sigma$ of the wormhole becomes a secondary black hole scalar hair, which is not independent, but a specific function of the mass. Furthermore, we find that black holes with $\Sigma >M$, $\Sigma = M$ and $\Sigma <M$ can all arise, as illustrated in Fig.~\ref{Fig20032}. This implies that the long-range force between two identical such black holes can be repulsive, zero and attractive. A further interesting phenomenon is that black holes with negative mass can also arise. The fact that phantom matter can give rise to black holes with negative mass is not common, but has been noted in literature. Our finding suggests that black holes with negative mass or black holes that repulse each other may be inevitable in theories accepting exotic matter for traversable wormholes.

\section*{Acknowledgement}

This work is supported in part by the National Natural Science Foundation of China (NSFC) grants No.~12575061, No.~12375052 and No.~11935009, and by the Tianjin University Self-Innovation Fund Extreme Basic Research Project Grant No.~2025XJ21-0007, and also by the Tianjin University Graduate Liberal Arts and Sciences Innovation Award Program (2023) No.~B1-2023-005.

\appendix

\section*{Appendix}

\section{Graphs of numerical results}

In this appendix, we give all the graphs that illustrate the black hole numerical construction and the properties discussed in section 4.

\begin{figure}[ht]
	\centering
	\includegraphics[width=0.45\textwidth]{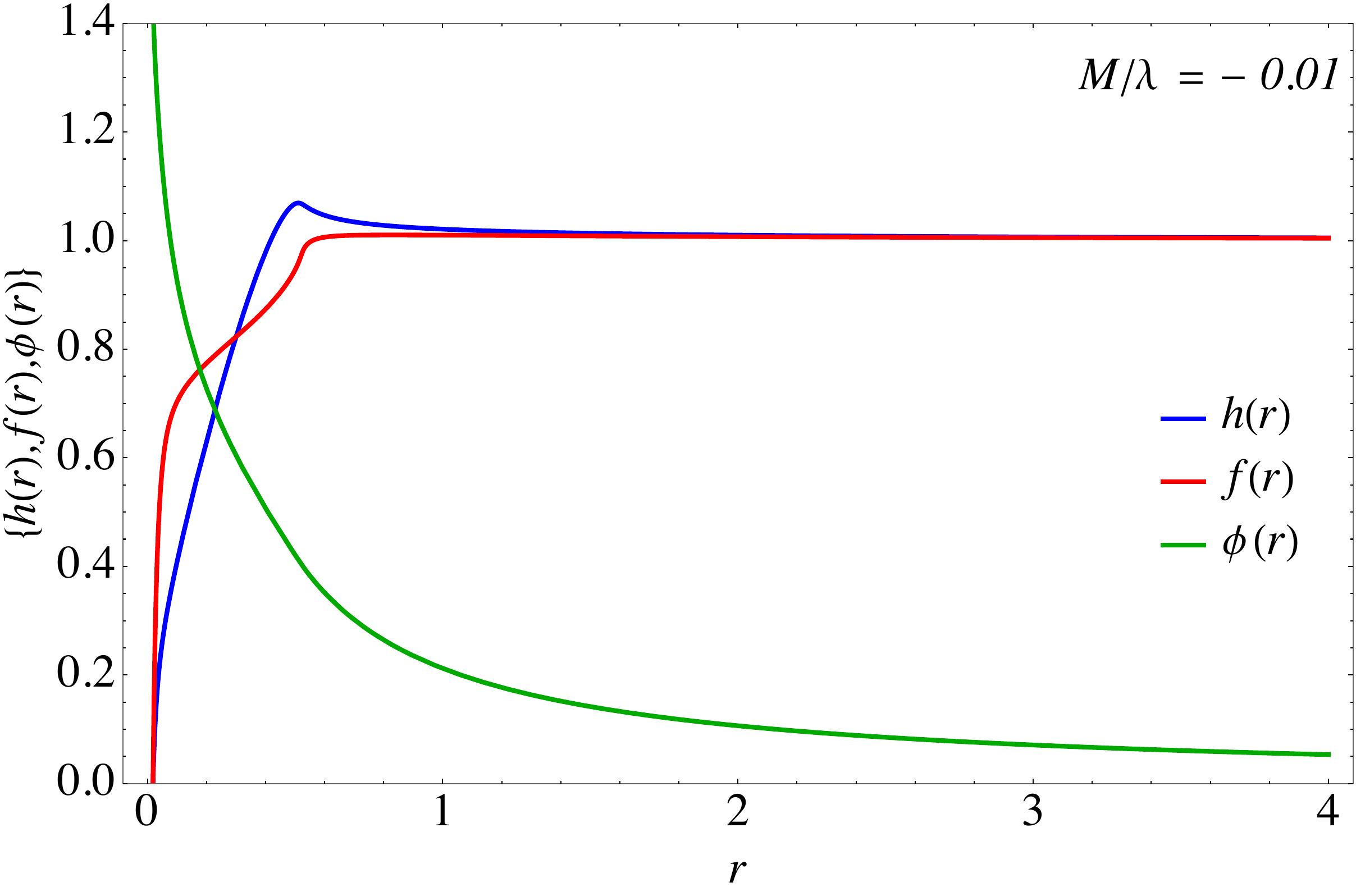}
	\qquad
	\includegraphics[width=0.45\textwidth]{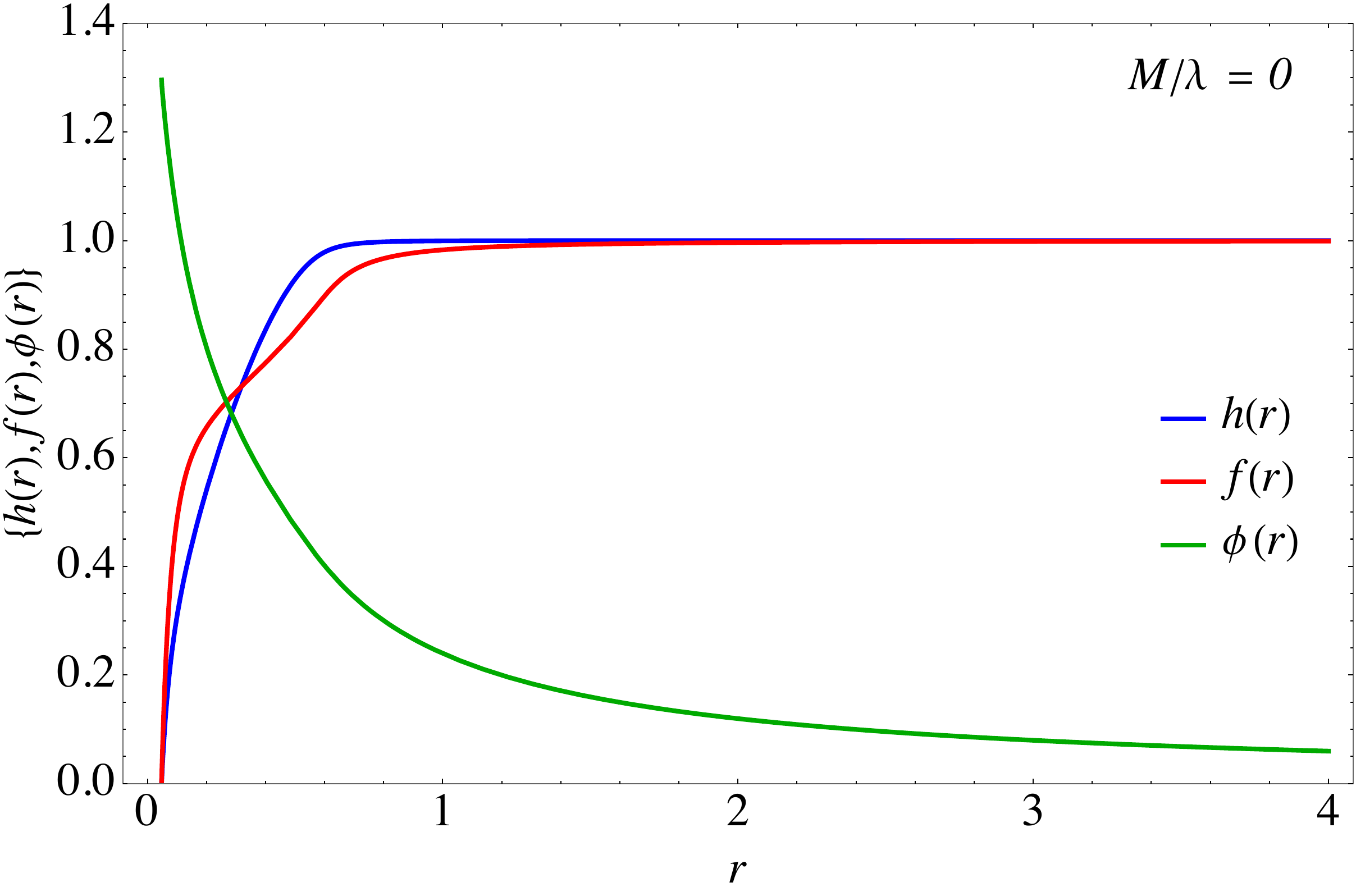}
	\qquad
	\includegraphics[width=0.45\textwidth]{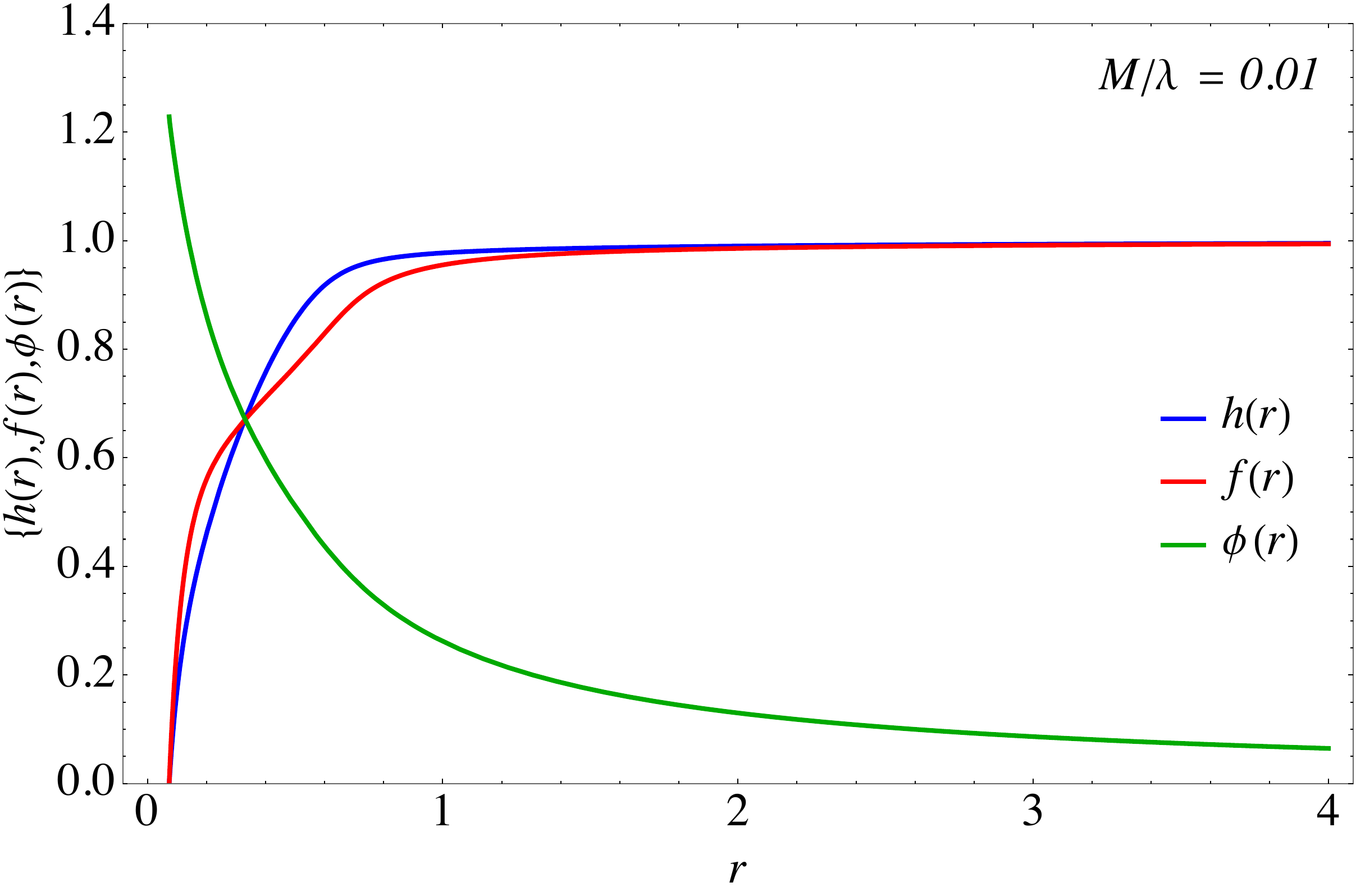}
	\qquad
	\includegraphics[width=0.45\textwidth]{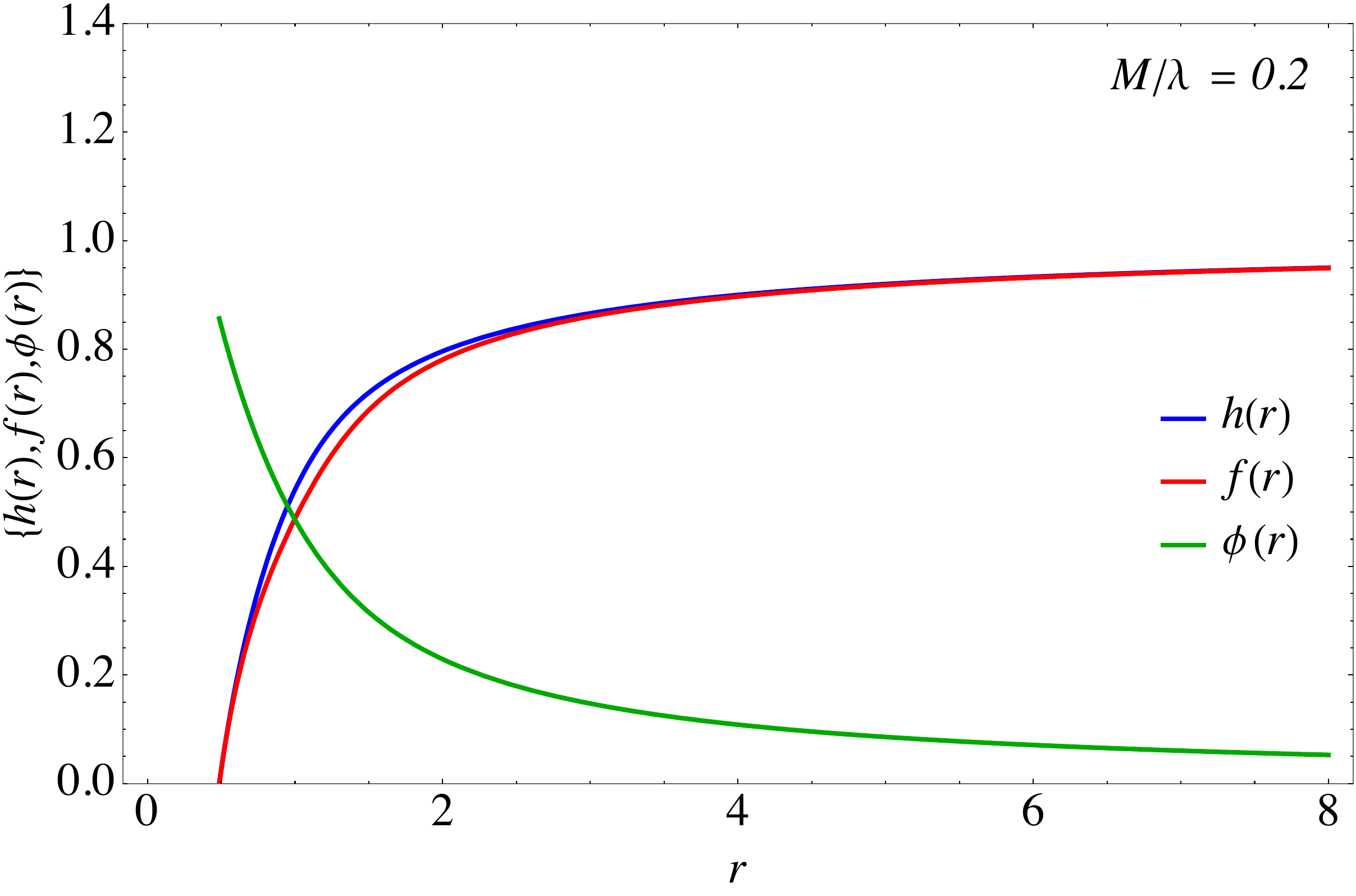}
	\caption{Radial profiles of the metric functions $h(r)$, $f(r)$ and scalar field $\phi(r)$ for black hole solutions with different mass $M/\lambda=-0.01, 0, 0.01$, and $0.2$. We see that influence of the scalar field becomes lesser as the mass increases.}
	\label{Fig20021}
\end{figure}

\begin{figure}[ht]	
	\centering
	\includegraphics[width=0.45\textwidth]{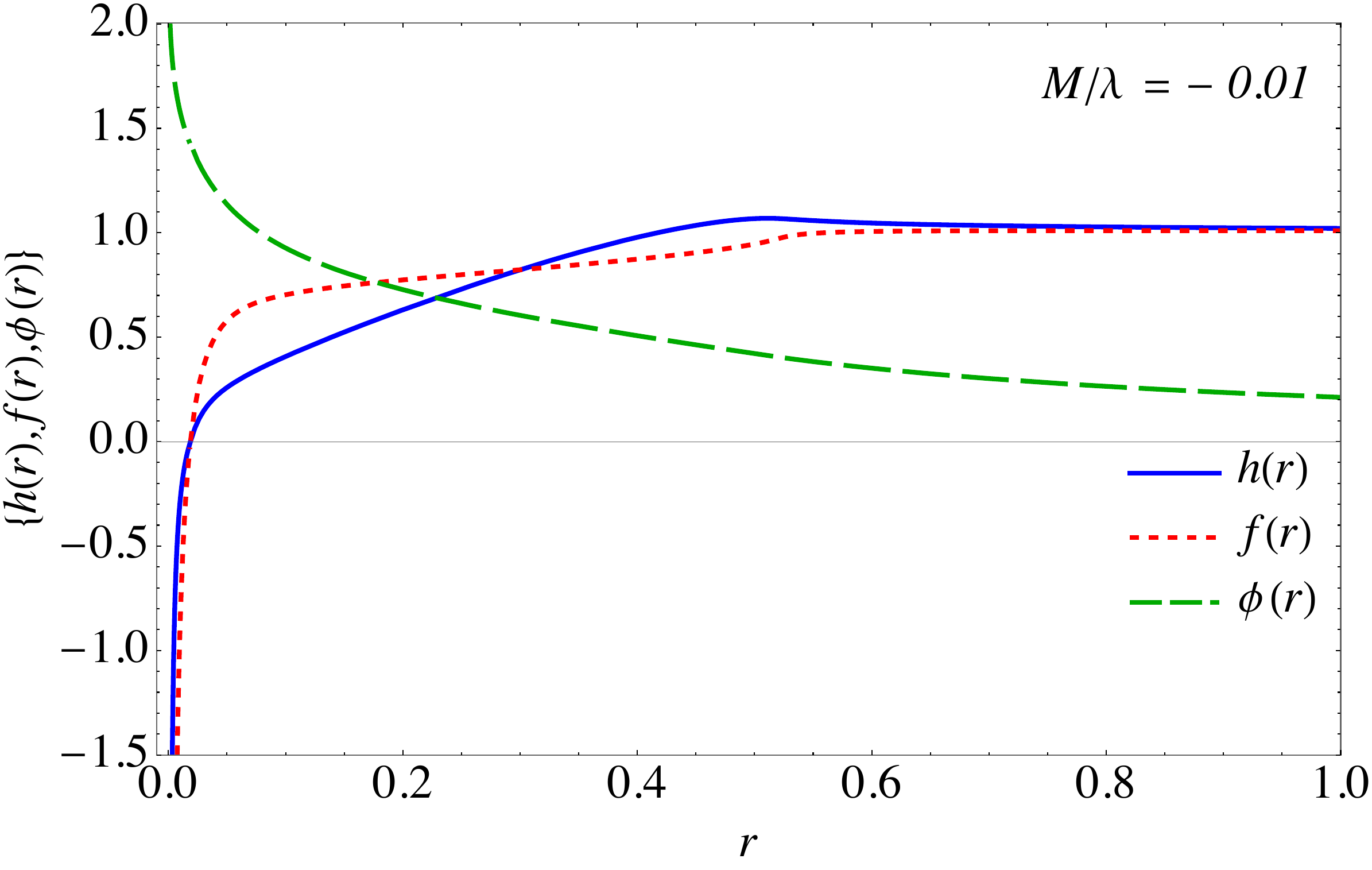}
	\qquad
	\includegraphics[width=0.45\textwidth]{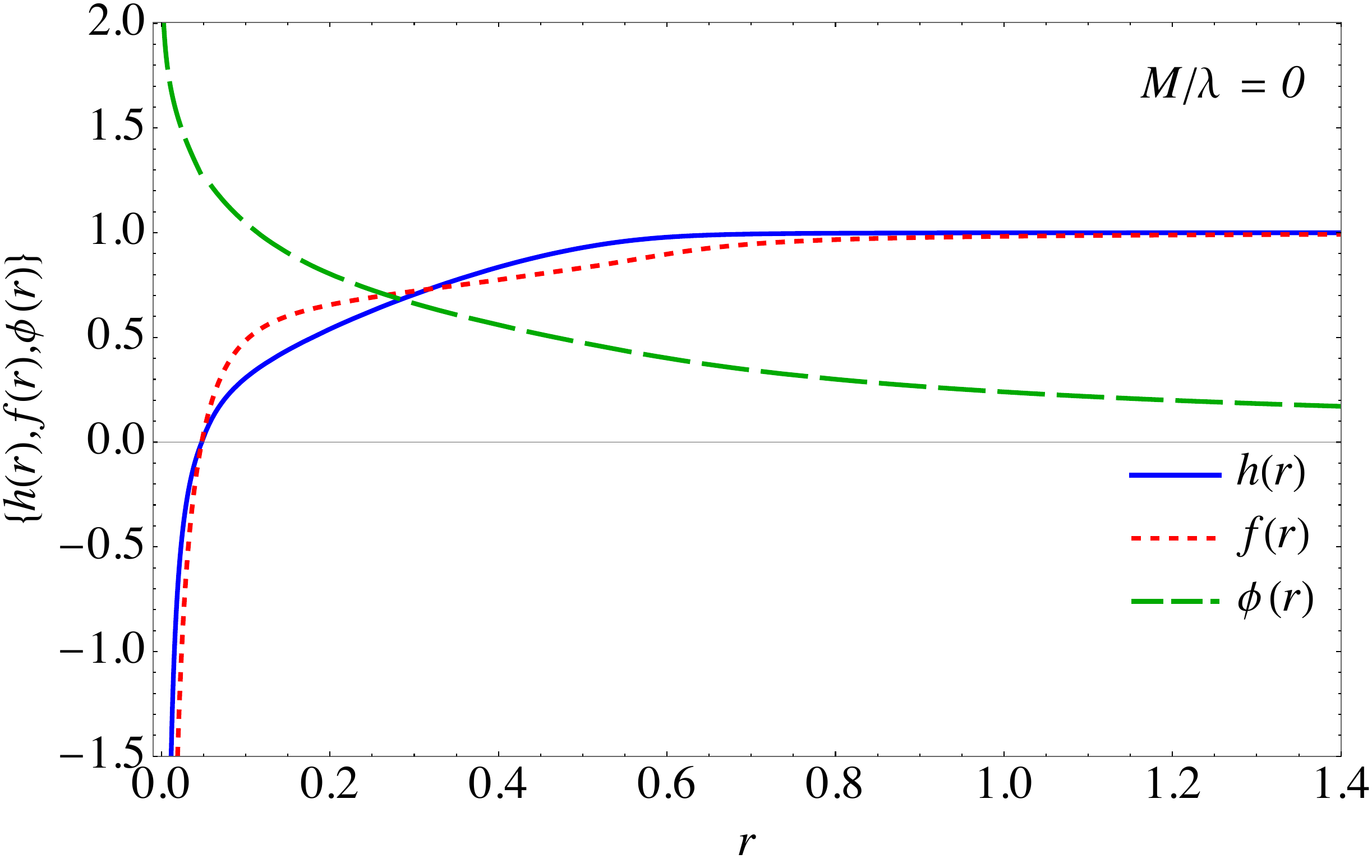}
	\qquad
	\includegraphics[width=0.45\textwidth]{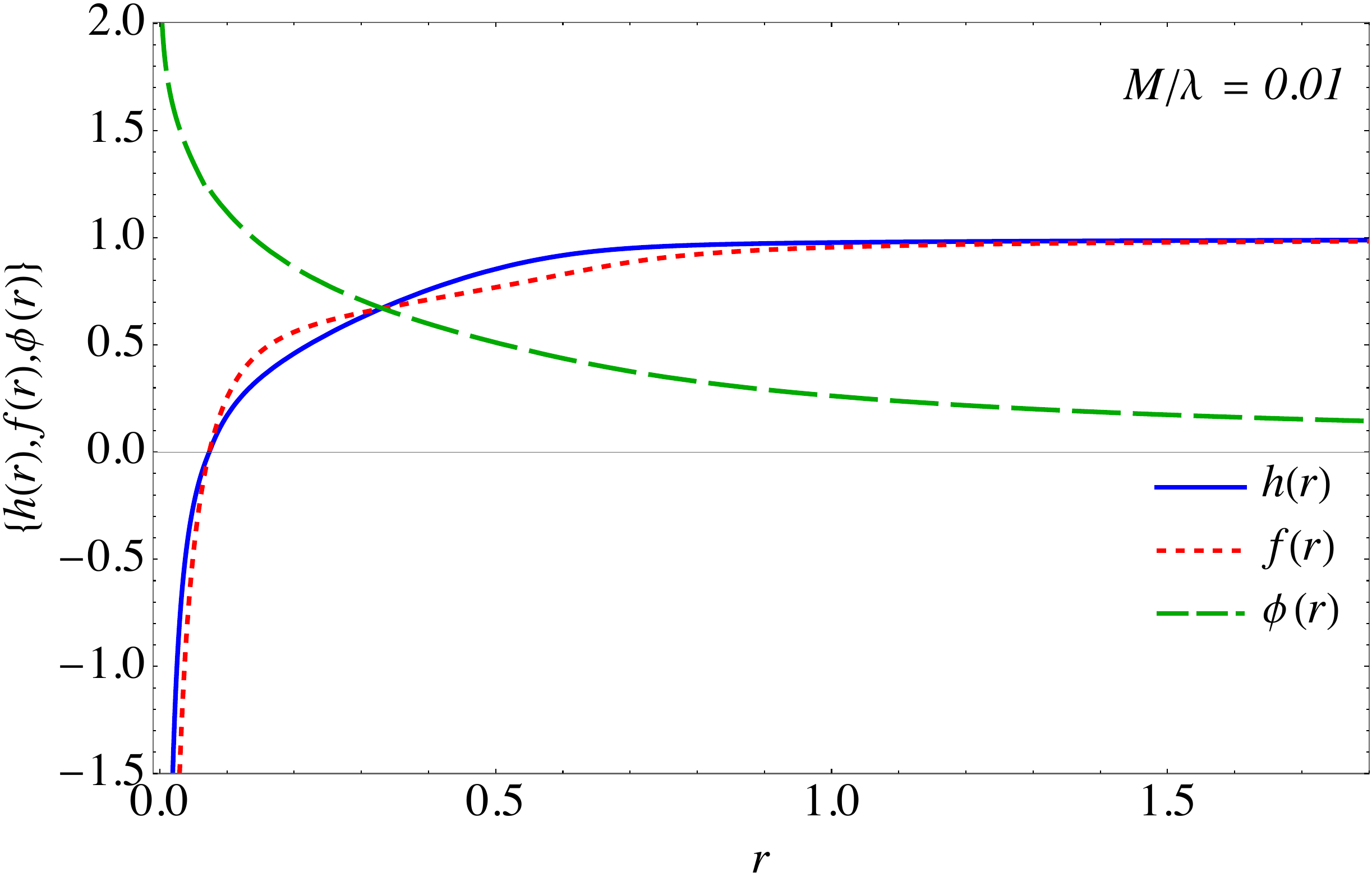}
	\qquad
	\includegraphics[width=0.45\textwidth]{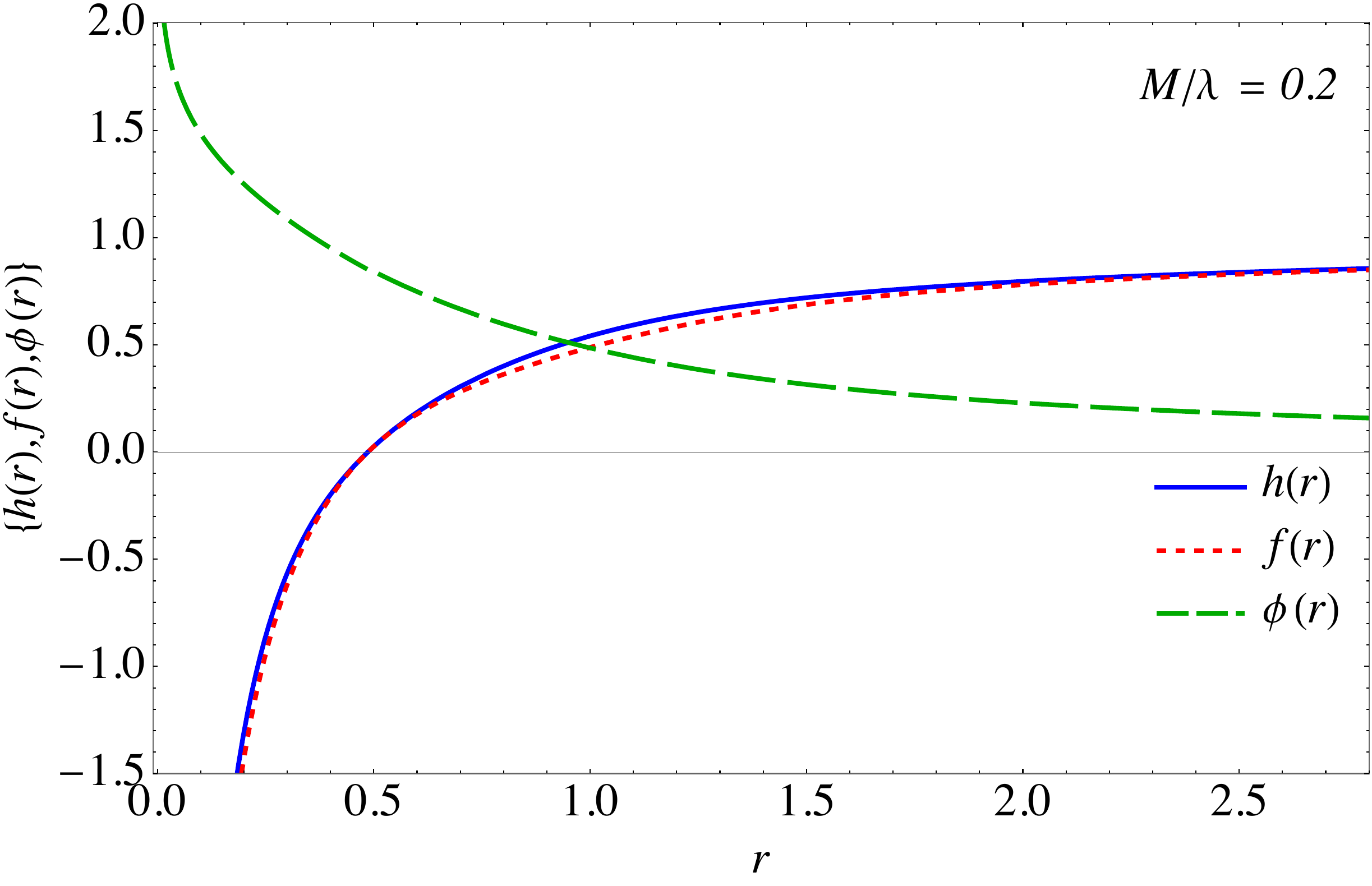}
	\caption{Interior structure of black hole solutions showing the divergent behavior of metric functions $h(r)$ and $f(r)$ near the singularity at $r=0$. The plots correspond to four distinct mass values:  $M/\lambda=-0.01, 0, 0.01$, and $0.2$.}
	\label{Fig200211}
\end{figure}

\begin{figure}[ht]	
	\centering
	\includegraphics[width=0.31\textwidth]{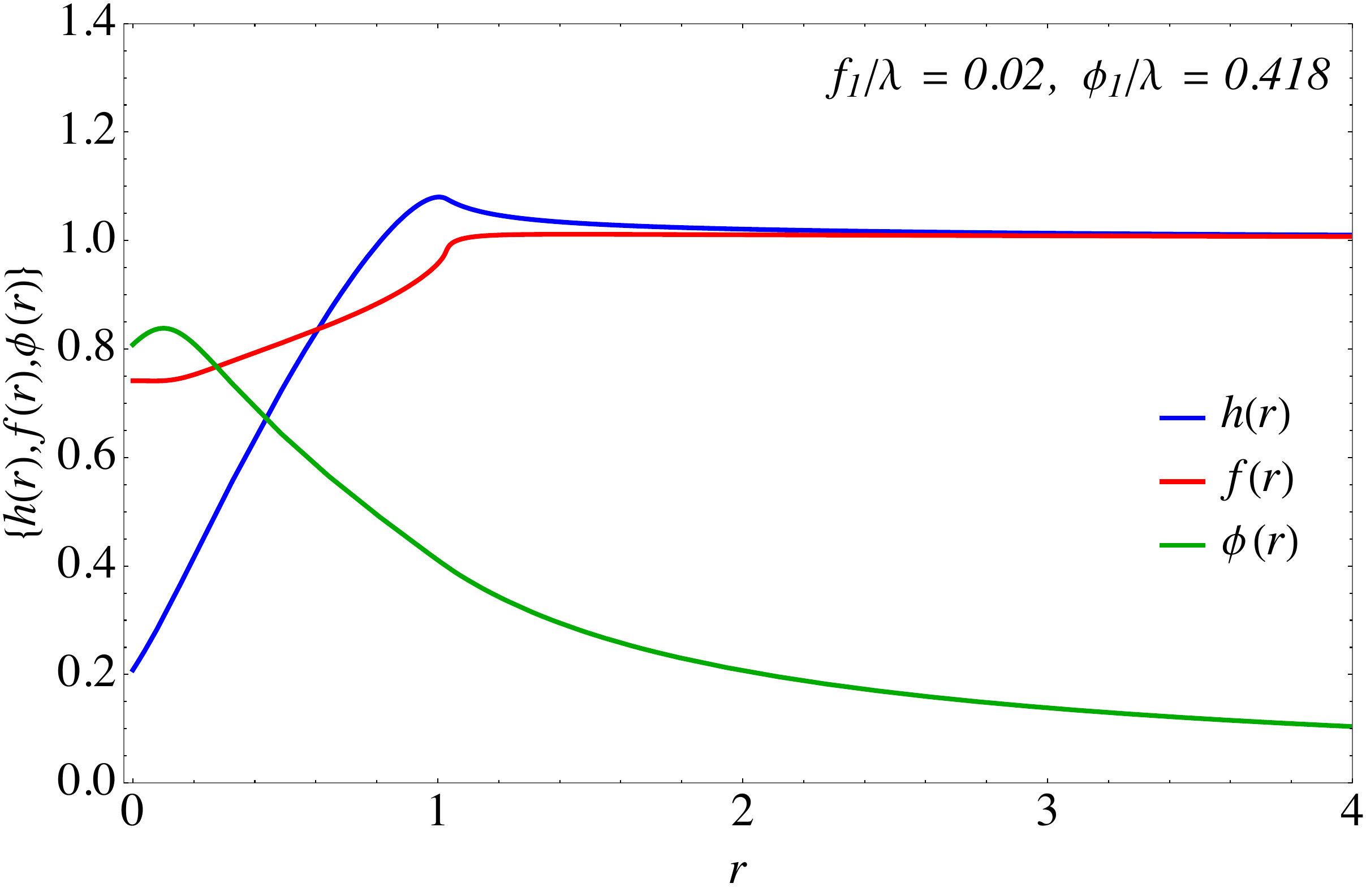}
	\quad
	\includegraphics[width=0.31\textwidth]{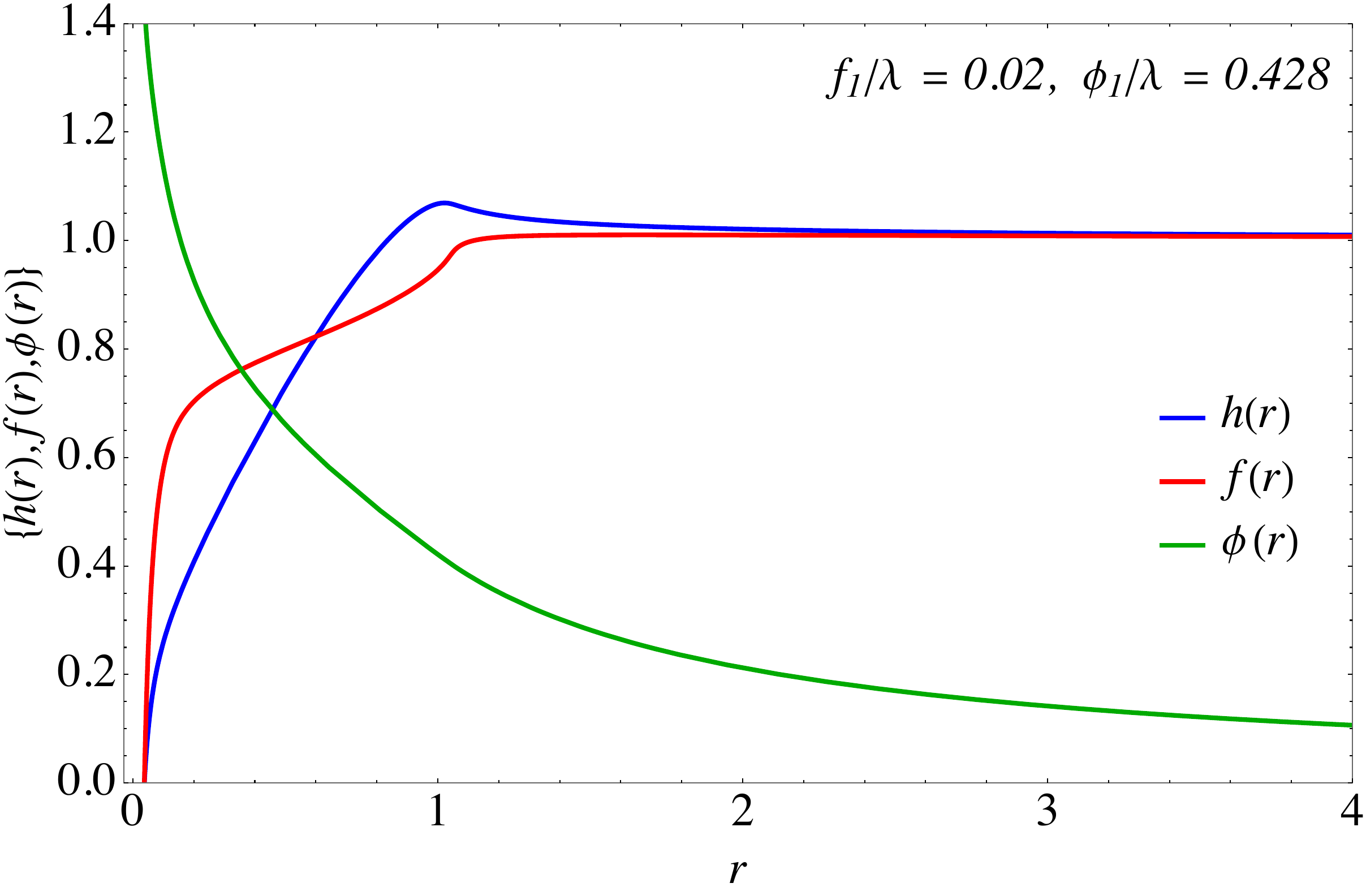}
	\quad
	\includegraphics[width=0.31\textwidth]{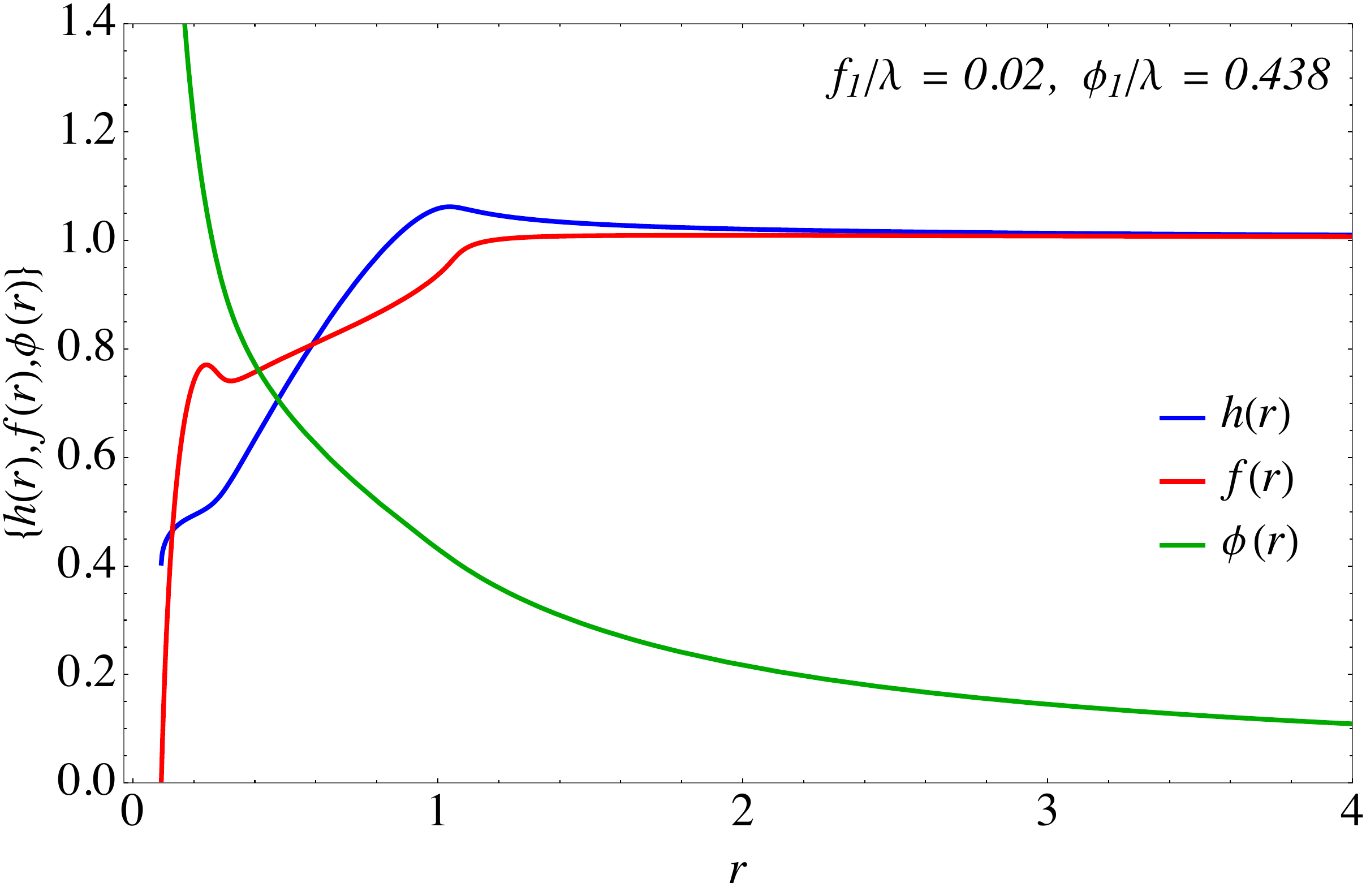}
	\caption{Negative mass solutions for fixed $f_1 = 0.02$ shows the transition between different spacetime configurations as $\phi_1$ varies. The middle curve gives a black hole, flanked by ``almost regular'' horizonless solutions and wormhole solutions.}
	\label{Fig20038}
\end{figure}

\begin{figure}[ht]	
	\centering
	\includegraphics[width=0.45\textwidth]{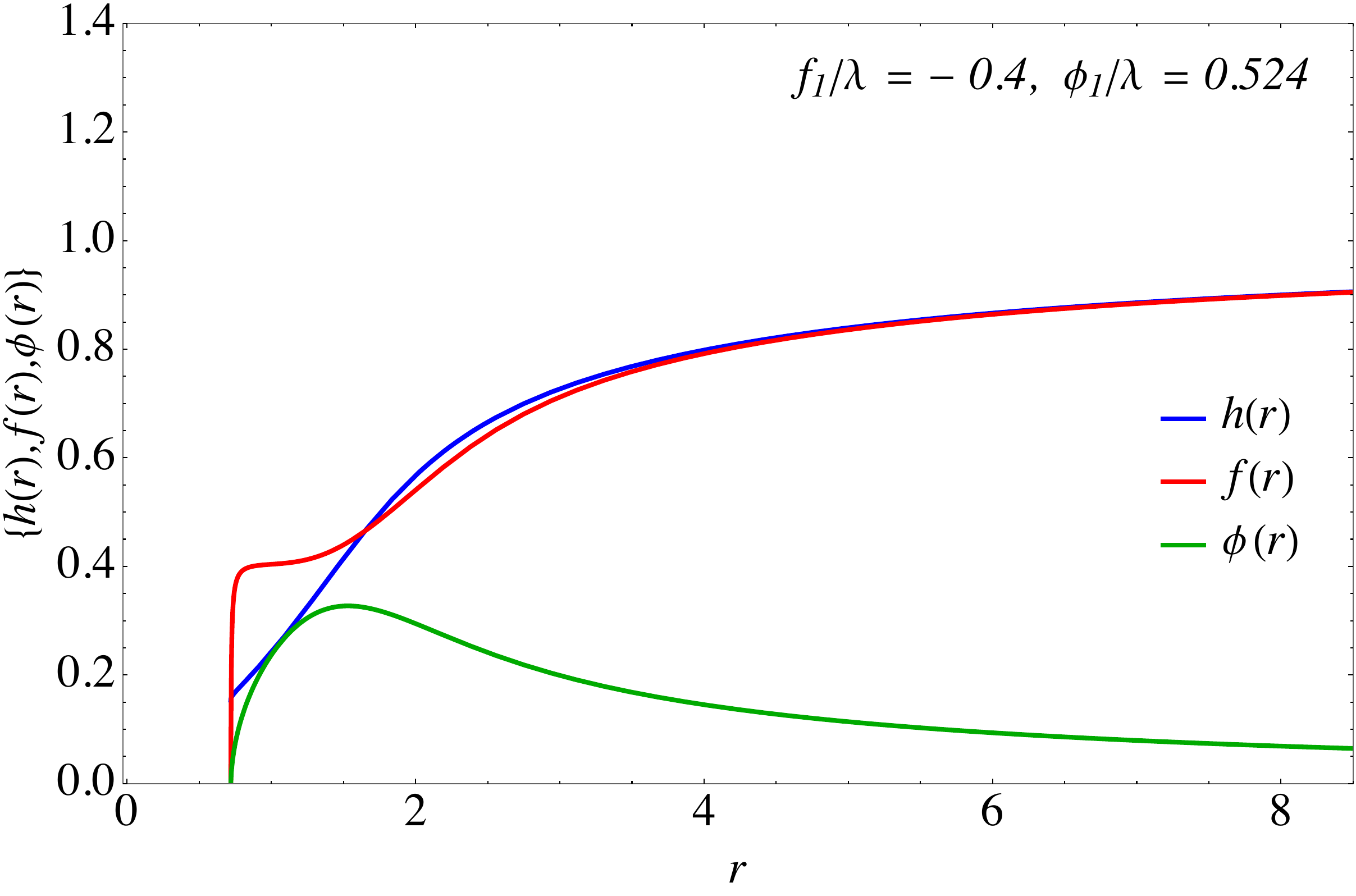}
	\qquad
	\includegraphics[width=0.45\textwidth]{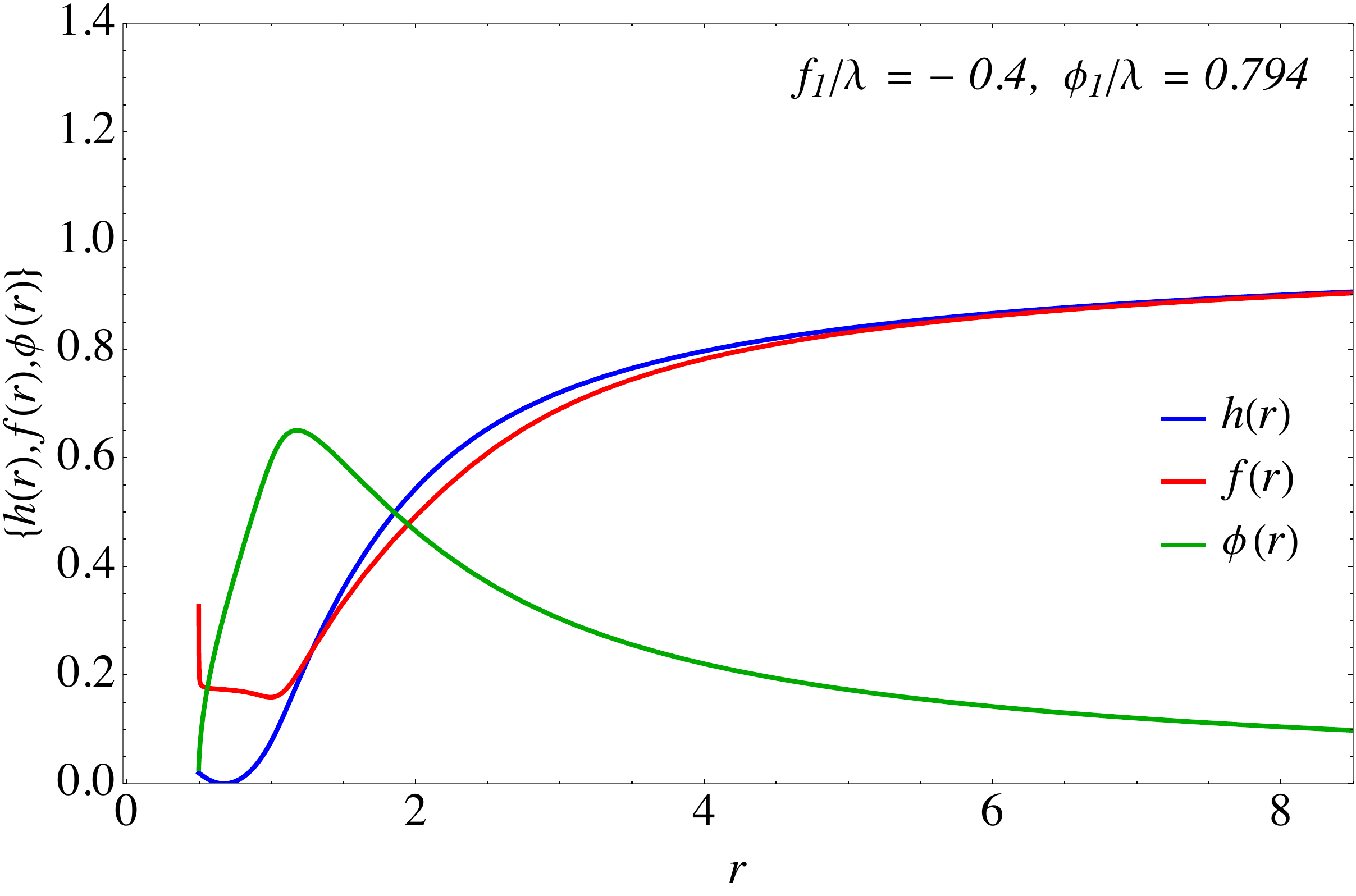}
	\qquad
	\includegraphics[width=0.45\textwidth]{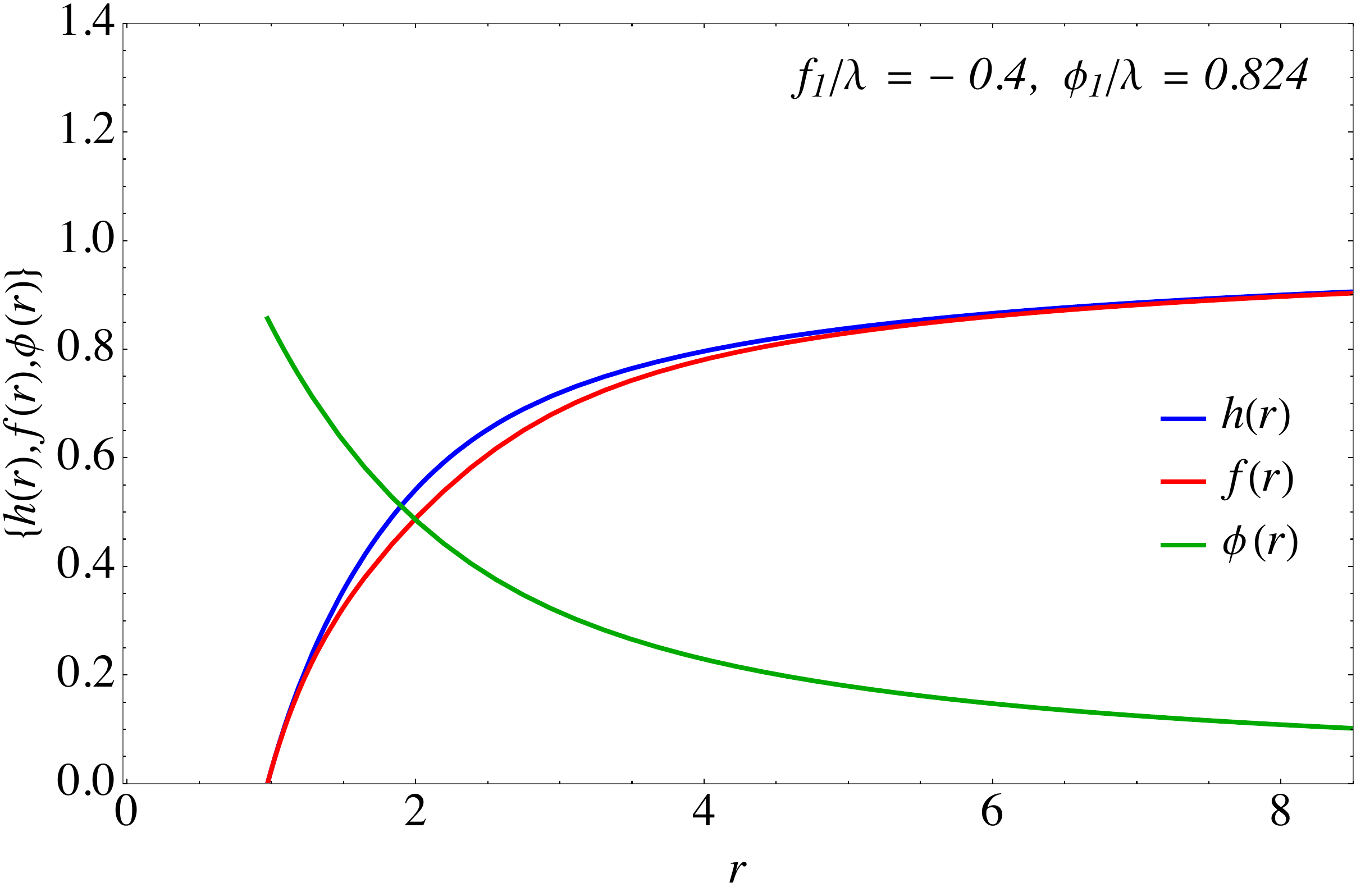}
	\qquad
	\includegraphics[width=0.45\textwidth]{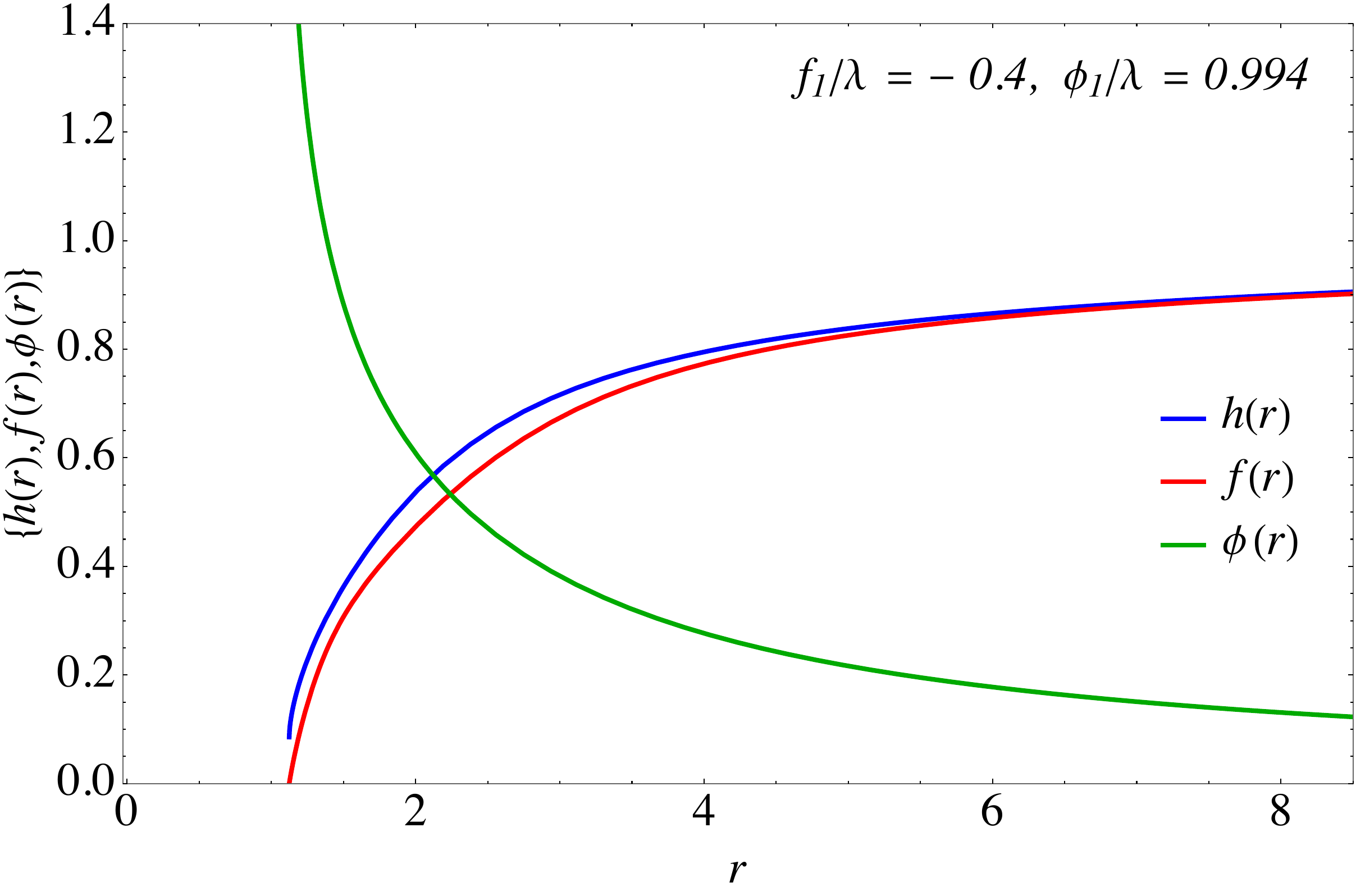}
	\caption{Rich solution space for $f_1 = -0.4$ showing the emergence of different spacetime geometries depending on the scalar field parameter $\phi_1$.}
	\label{Fig20039}
\end{figure}

\begin{figure}[ht]	
	\centering
	\includegraphics[width=0.75\textwidth]{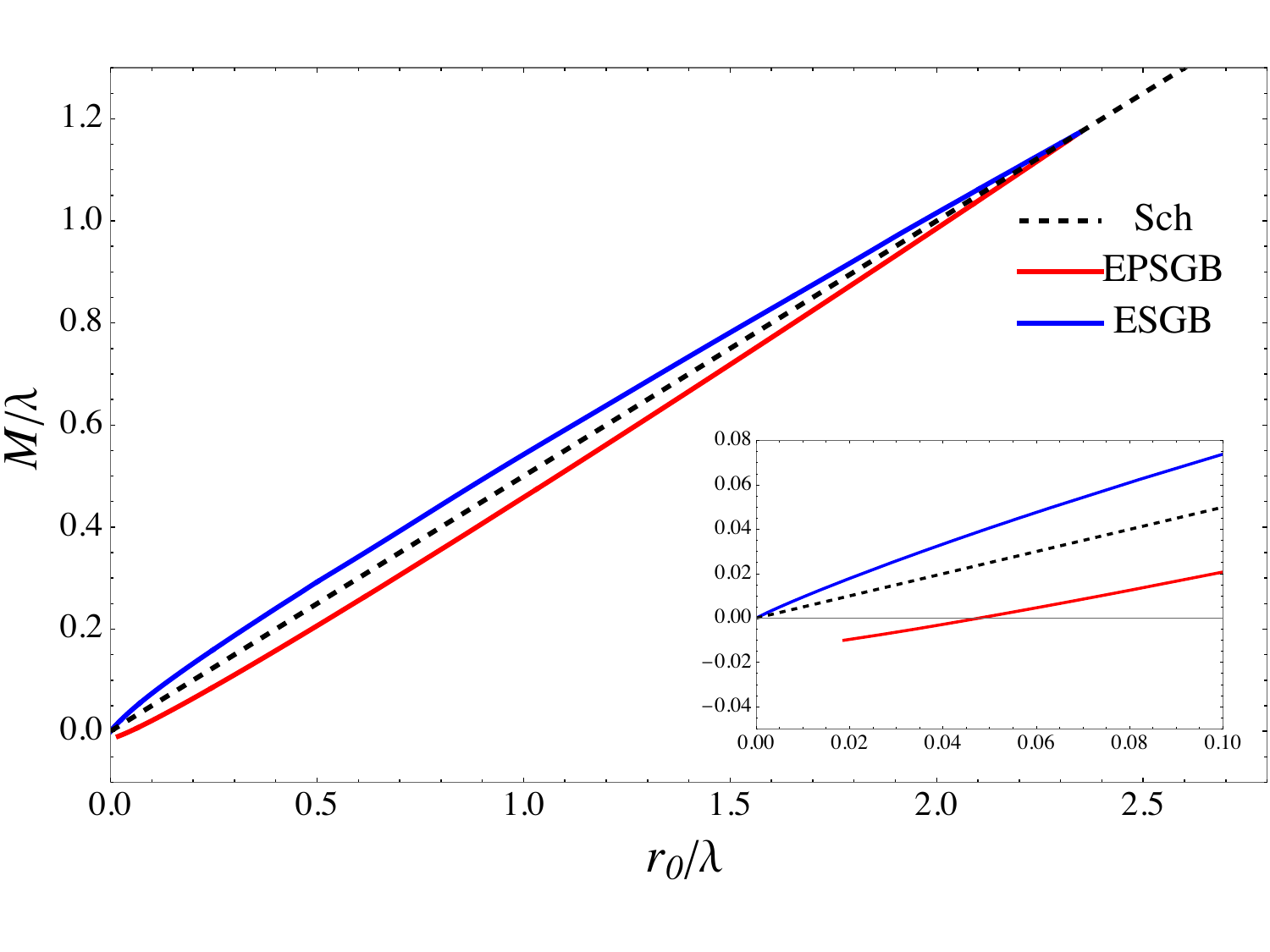}
	\caption{Mass-horizon radius ($ M-r_0 $) relation for scalarized black holes. Dashed curve: Schwarzschild solution; red solid curve: EPSGB theory ($ \varepsilon=-1 $); blue solid curve: ESGB theory ($ \varepsilon=1 $). The critical mass $ M_{\text{max}} $ (black dot) is shared by both theories, but EPSGB solutions (red) have $ M < M_{\text{Schwarzschild}} $, while ESGB solutions (blue) exhibit $ M > M_{\text{Schwarzschild}} $. The EPSGB branch also admits negative masses, with a minimum $ M_{\text{min}} \approx -0.01 $ at $ r_0 \approx 0.0188 $.}
	\label{Fig20022}
\end{figure}

\begin{figure}[ht]	
	\centering
	\includegraphics[width=0.76\textwidth]{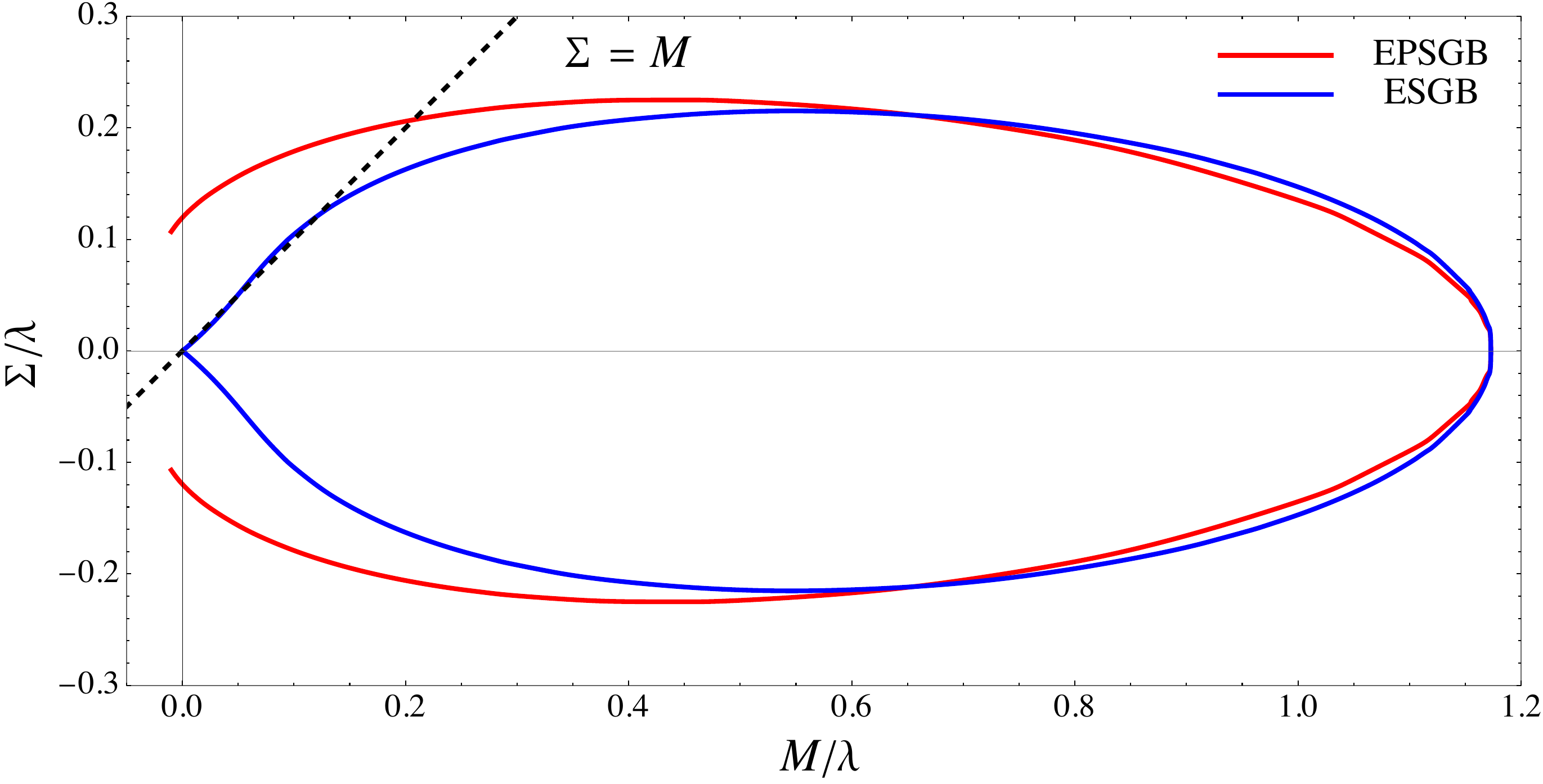}
	\caption{Scalar charge $\Sigma$ as a function of mass $M$. Red solid curve: EPSGB solutions; blue solid curve: ESGB solutions; Dashed line: $\Sigma = M$. The ESGB branch forms a closed loop, while the EPSGB curve remains open-ended at low masses (left side). Both theories share the same right endpoint at $M_{\text{max}} \approx 1.173$, which is $M_{\rm cr}$ mentioned in the introduction.}
	\label{Fig20032}
\end{figure}

\begin{figure}[ht]	
	\centering
	\includegraphics[width=0.45\textwidth]{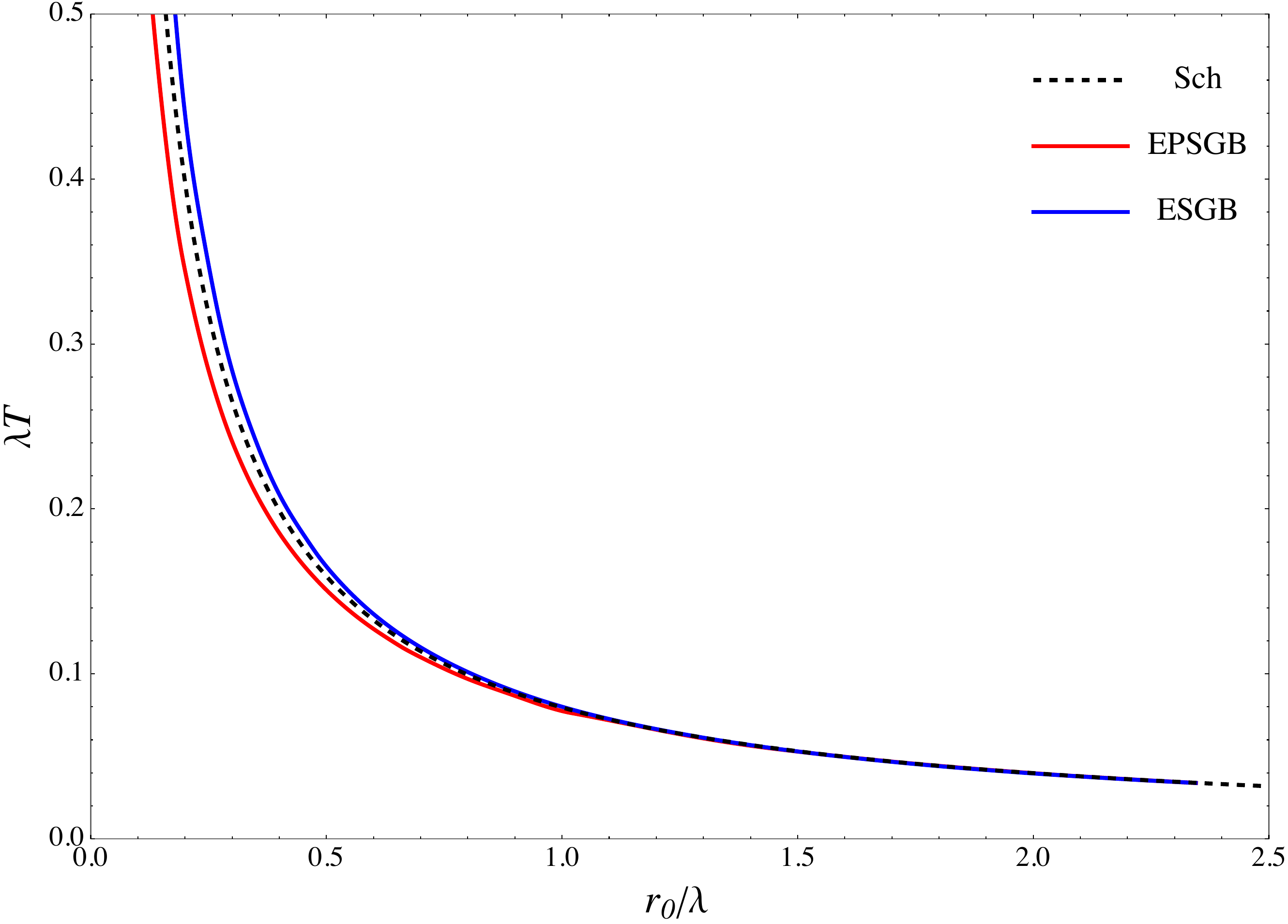}\ \includegraphics[width=0.45\textwidth]{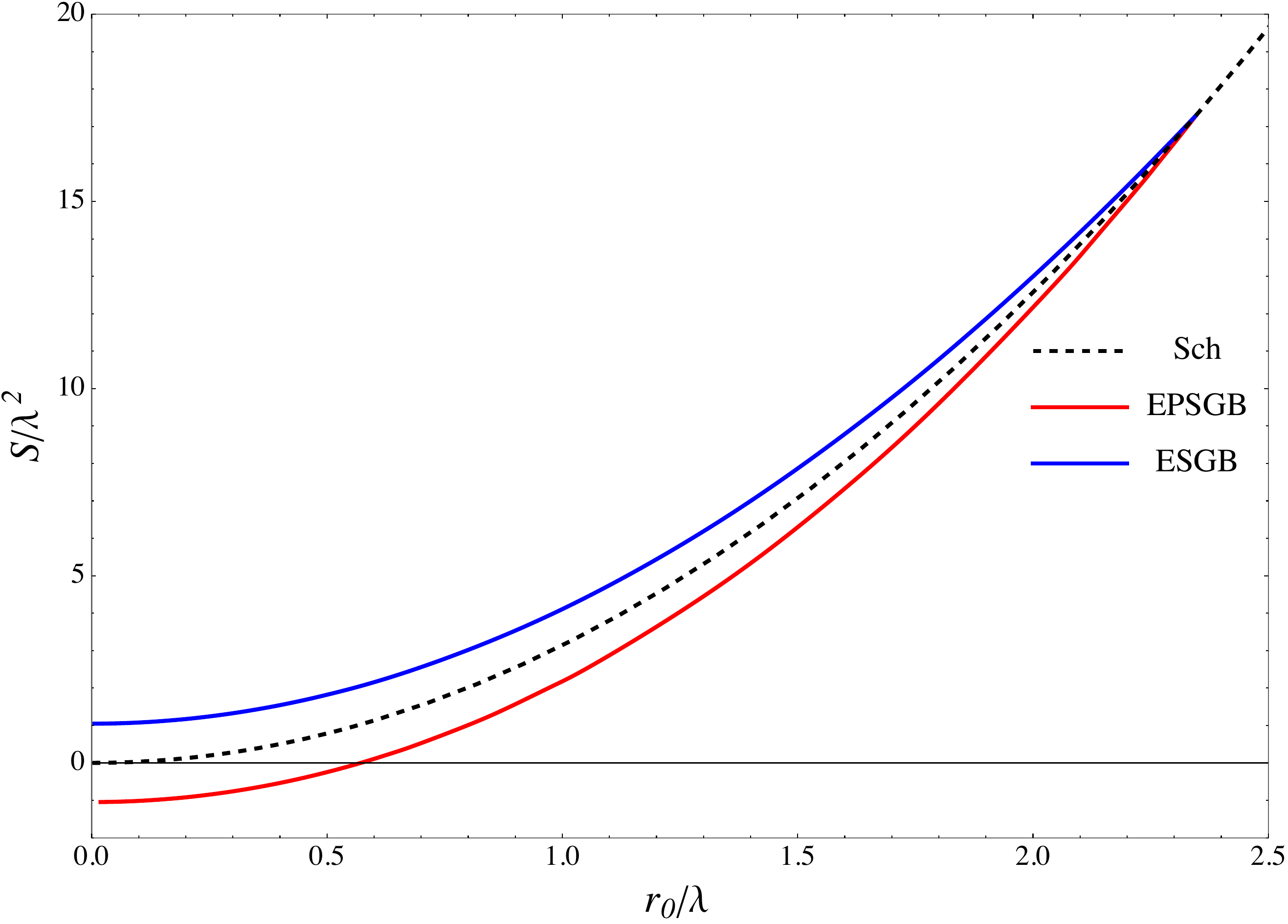}
	\caption{Temperature and entropy versus horizon radius. The scalarized solutions (solid curves) deviate from Schwarzschild (dashed) except at $r_0 = 2.345$, where both theories predict identical temperature and entropy.  The entropy becomes negative for $r_0 < 0.327$ (vertical dotted line)}
	\label{Fig20033}
\end{figure}

\begin{figure}[ht]	
	\centering
	\includegraphics[width=0.45\textwidth]{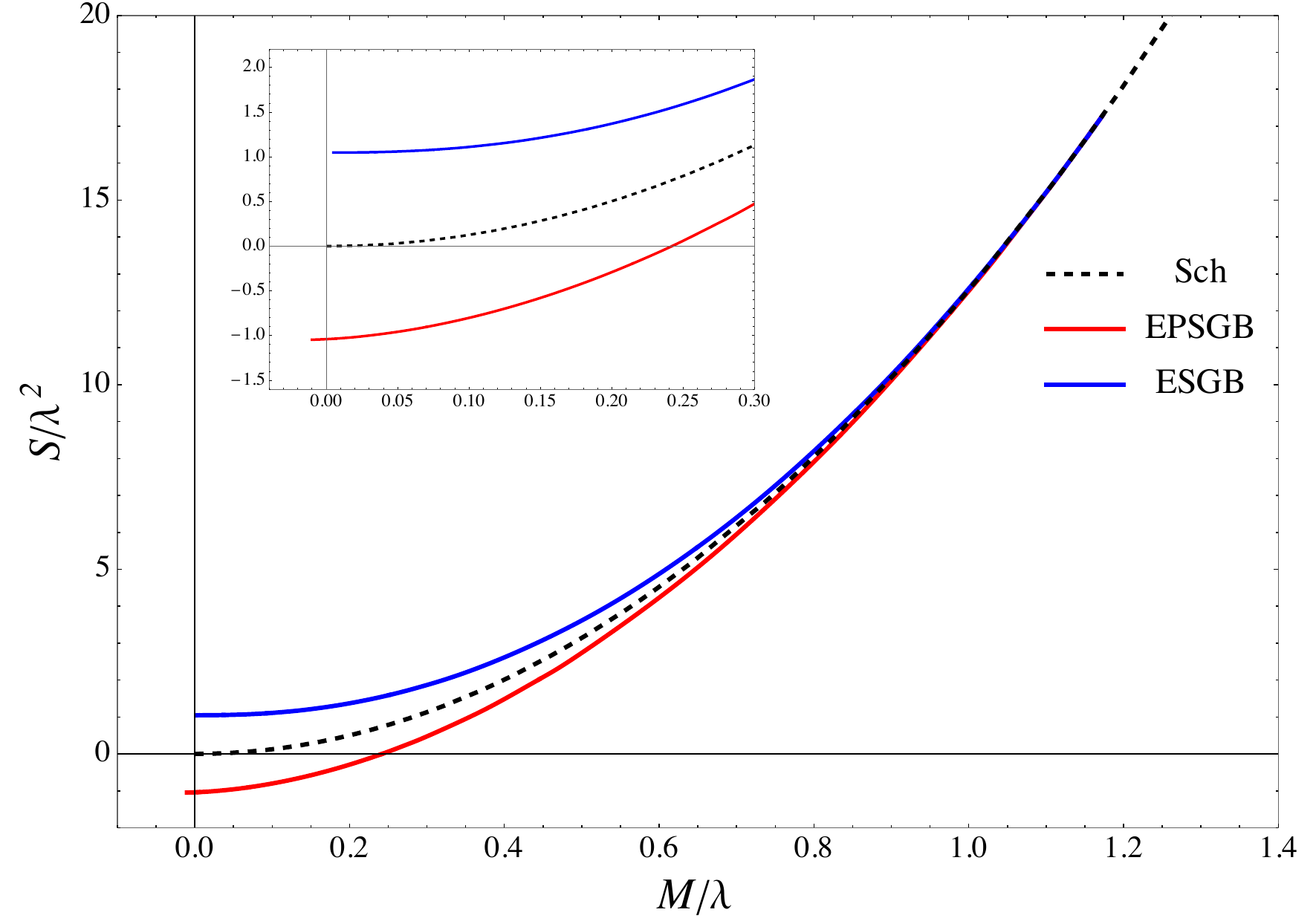}\
\includegraphics[width=0.45\textwidth]{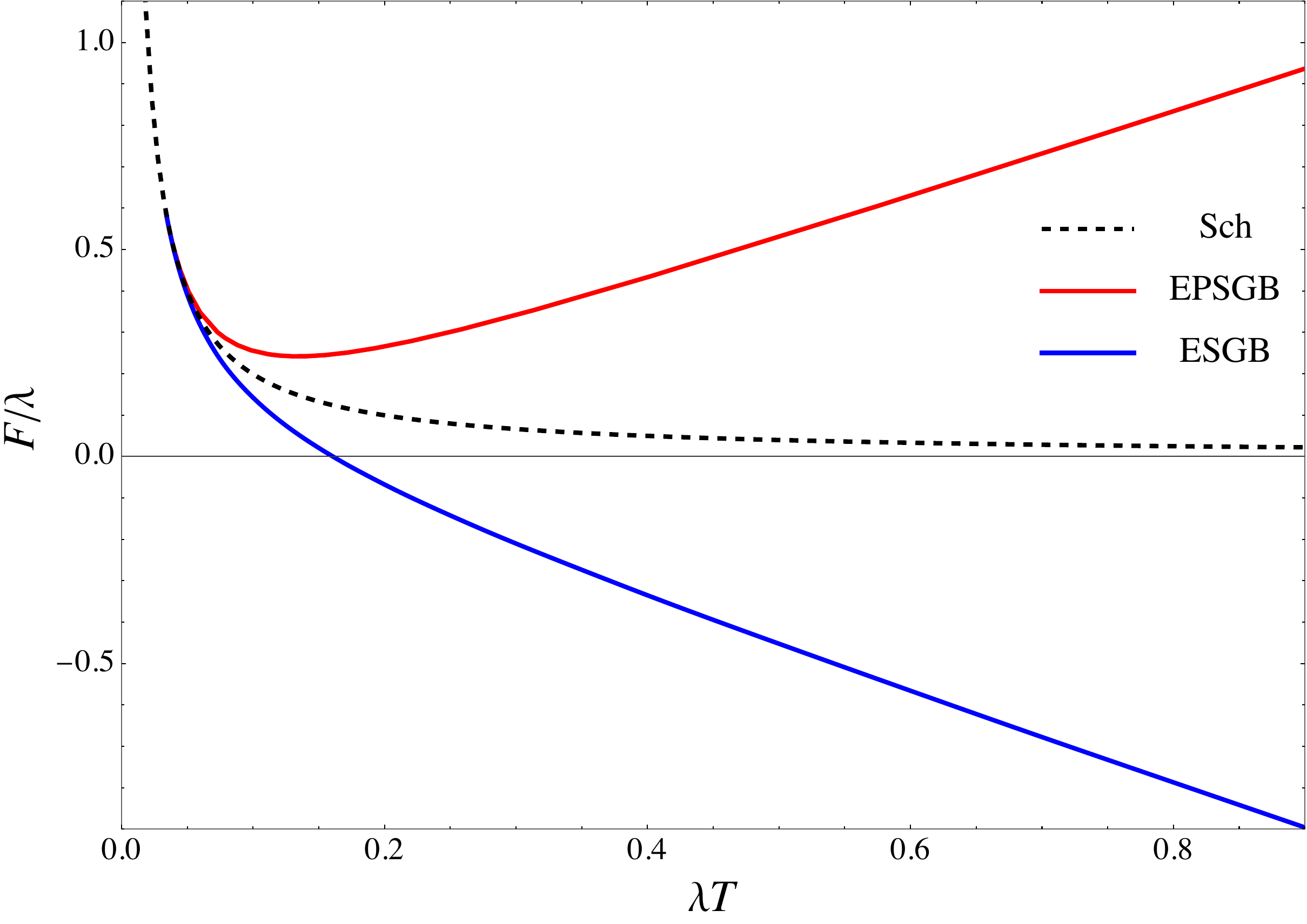}
	\caption{The left panel depicts the entropy as a function of mass. The EPSGB (ESGB) solutions (red curve (blue curve)) always have smaller (larger) entropy than Schwarzschild black holes (dashed black curve) at given  mass. The blue dotted curve shows the fitted relation \eqref{eq:S_fit}. The right-panel depicts the free-energy versus temperature. EPSGB black holes (red curve) exhibit higher free energy than Schwarzschild (black dashed) at all temperatures, whilst the ESGB black holes (blue curve) has lower free-energy. ( The curves converge at $T_{\text{min}}$, corresponding to the maximum horizon radius.}
	\label{Fig200362}
\end{figure}

\begin{figure}[ht]	
	\centering
	\includegraphics[width=0.71\textwidth]{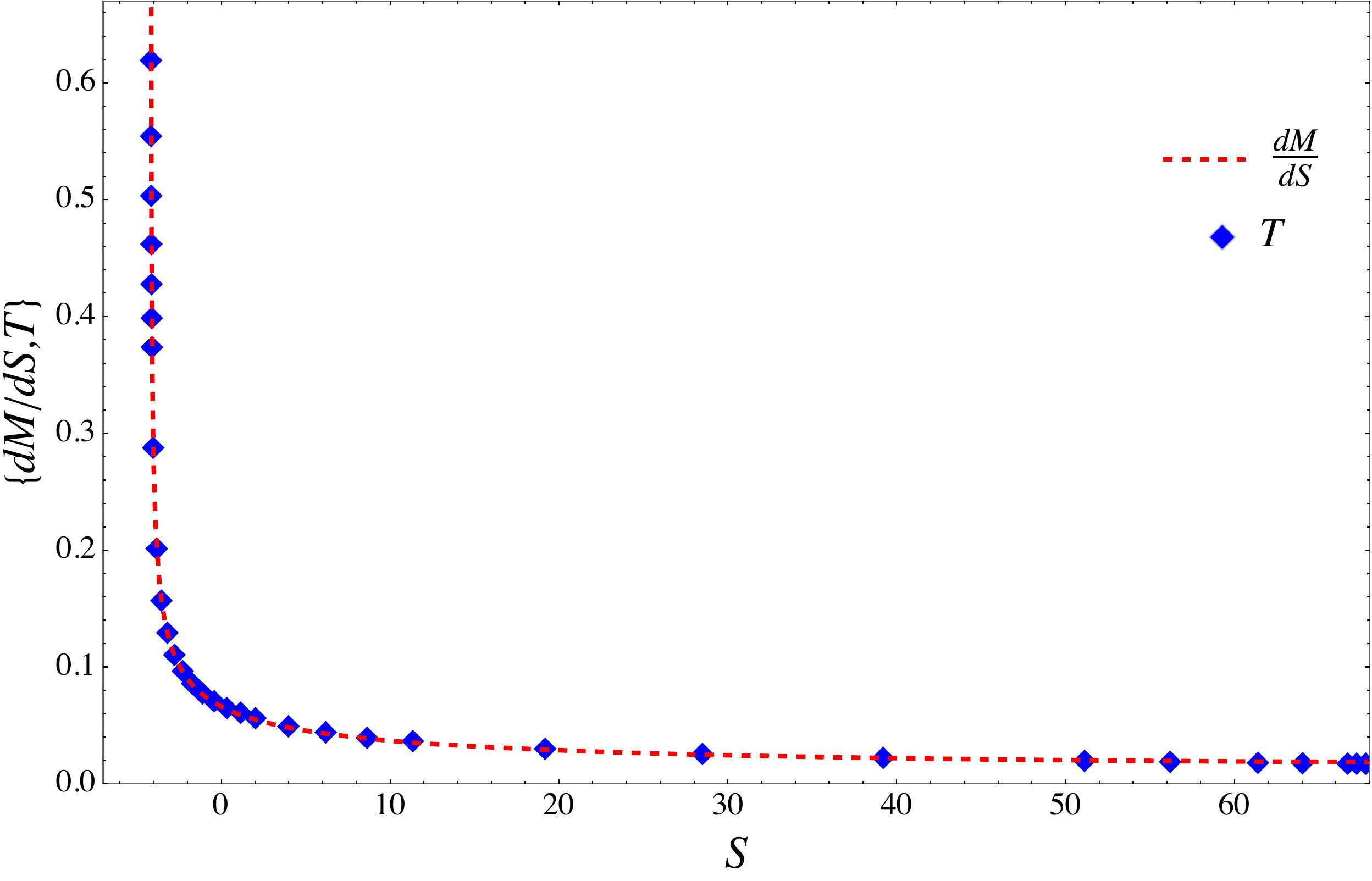}
	\caption{Verification of $\text{d} M / \text{d} S = T $. The red curve represents $\text{d} M / \text{d} S$ obtained by differentiating the interpolating function $M(S)$. The blue data points correspond to the temperature $T$.}
	\label{Fig20037}
\end{figure}

\end{document}